\documentclass[11pt]{article}

\usepackage{jheppub} 

\usepackage{graphicx}
\usepackage{amsmath, amssymb}

\newcommand{\be}{\begin{eqnarray}}
\newcommand{\ee}{\end{eqnarray}}
\newcommand{\bea}{\begin{eqnarray}}
\newcommand{\eea}{\end{eqnarray}}

\newcommand{\bn}{\begin{enumerate}}
\newcommand{\en}{\end{enumerate}}

\def\IZ{\mathbb{Z}}

\def\half{\frac{1}{2}}

\def\goto{\rightarrow}

\title{A freely generated  ring for ${\cal N}=1$ models in class $S_k$}

\preprint{}

\author{Shlomo S. Razamat,}

  \author{Evyatar Sabag}

\affiliation{Department of Physics, Technion, Haifa, 32000, Israel}

\emailAdd{razamat@physics.technion.ac.il, sevyatar@campus.technion.ac.il}

\abstract
{We study 4d ${\cal N}=1$ supersymmetric theories of class ${\cal S}_k$, obtained from flux compactifications on a Riemann surface of 6d $(1,0)$ conformal theories describing the low energy physics  on a stack of M5 branes probing a ${\mathbb Z}_k$ singularity. We conjecture that the protected spectrum of class ${\cal S}_k$ theories contains a freely generated  ring, generalizing the Coulomb branch of the ${\cal N}=2$ theories.
We derive this by examining a limit of the supersymmetric index of 4d ${\cal N}=1$ class ${\cal S}_k$ theories. The limit generalizes the Coulomb limit of ${\cal N}=2$ theories, which coincides with the case of $k=1$ for a particular choice of flux.  
We conjecture a general simple formula for the index in the aforementioned limit.
}

\begin{document} 

\maketitle
\flushbottom

\section{Introduction and Summary}
\label{S:intro}

Compactifying $(1,0)$ six dimensional theories to four dimensions on a Riemann surface with appropriate twists gives an ${\cal N}=1$ supersymmetric quantum field theory in four dimensions.
Since the six dimensional models lack semi-classical description at the conformal point, it is challenging to understand what is the dependence of a specific four dimensional theory on the compactification choices.
Often if one looks for semi-classical descriptions of the four dimensional theories, insisting on having all the symmetries manifest, no description can be found. However, some information about the four dimensional models can be rather easily deduced, mainly by the use of anomalies and symmetries as well as the spectrum of some simple protected operators. One can also deduce in many cases dualities relating different models obtained in the compactification.

In recent years, most of the research on compactifications of six dimensional theories has been focused on the $(2,0)$ theory of type $A_{N-1}$, which is the model on a stack of $N$ M5 branes \cite{Gaiotto:2009we}. Compactifying to four dimensions with a particular twist leads to models with extended supersymmetry.
This in particular allows for additional robust properties in these theories to be extractable. Most notably one can define the notion of a Coulomb branch, which is believed to be a freely generated ring of chiral operators, and compute the Seiberg-Witten curve associated to it \cite{Seiberg:1994rs}. In fact, studying the curves was very instrumental \cite{Gaiotto:2009we} in the understanding of the precise map between compactifications and the four dimensional theories. 

When we reduce the amount of supersymmetry to the minimal one,  there is no longer a natural definition of the Coulomb branch prescribed just by supersymmetry. An alternative definition is that the Coulomb branch is the part of the moduli space on which the gauge groups are broken to abelian factors. For particular examples of models, one can try and extend the success of the aformentioned extended supersymmetry case \cite{Intriligator:1994sm} to set-ups with less supersymmetry, though the result is not  expected to be as sensational. This approach was taken recently for the compactifications of M5 branes probing ${\mathbb Z}_k$ singularity \cite{Coman:2015bqq,  Mitev:2017jqj, Bourton:2017pee}, in order to try and extend the AGT correspondence \cite{Alday:2009aq}. 

Yet another approach one can take is to generalize the notion of the Coulomb branch to be defined by a freely generated ring, or sub-ring, of protected operators. Such a ring might exist or might be trivial for a given model.
This is the route we will follow in the current note. We will consider very specific four dimensional models mentioned in the previous paragraph. These are obtained by compactifying the six dimensional theory living on a stack of M5 branes probing $A_{k-1}$ singularity on a Riemann surface with flux (class ${\cal S}_k$ theories of \cite{Gaiotto:2015usa}, see also \cite{Franco:2015jna, Hanany:2015pfa, Maruyoshi:2016caf,Razamat:2016dpl,Bah:2017gph,Yagi:2017hmj}). Such models have an $SU(k)^2\times U(1)_t$ symmetry, which for special choices of the singularity and number of branes might be enhanced. 
We will argue that compactifying with $U(1)_t$ flux $e>0$, results in a four dimensional theory that has a  freely generated ring of protected operators, as can be seen in a certain limit of the superconformal index.  We will characterize the properties of the operators appearing in the ring for any choice of compactification. In particular, we will give a prediction which will not depend on having a weakly coupled description of the model.   

On a technical level, the conjecture we make is possible due to the analysis of the supersymmetric index of theories for which the index is known and extrapolating the result to other cases.
The known instances are the class ${\cal S}_1$ with general $N$, denoted often as class ${\cal S}$,  the case of class ${\cal S}_2$ and $N=2$ \cite{Gaiotto:2015usa, Razamat:2016dpl}, and the case of general $k$ and $N$ taken on a torus with flux \cite{Bah:2017gph}. We then perform several checks of the conjecture exploiting the dualities implied by the compactifications. These dualities, similar to the Argyres-Seiberg duality of class ${\cal S}$ \cite{Argyres:2007cn}, can be derived by considering theories corresponding to spheres with punctures, with the use of puncture closing procedure initiated by giving vacuum expectation values \cite{Gaiotto:2015usa}. Assuming that the compactification surface can be split into a collection of spheres with three punctures in various ways leading to an equivalent fixed point theory, one can find predictions of dualities involving strongly coupled SCFTs. The dualities determine some of the properties of the new strongly coupled SCFTs, such as 't Hooft anomalies, and also constrain the indices. In particular in the limit we will consider they determine the index. We will in appendix B state two such new dualities following the procedure of \cite{Gaiotto:2015usa}.

We will conjecture that, at least for some range of the $U(1)_t$ symmetry flux, the index of all the models has a limit, analogous to the Coulomb limit of ${\cal N}=2$ theories \cite{Gadde:2011uv,Beem:2012yn}, which is well defined. 
In this limit the index can be interpreted as obtaining contributions only from products of a finite set of bosonic and fermionic operators with no non trivial relations. This is the freely generated ring.
Let us quote here the result for the generating function of operators in this ring when compactifying on a Riemann surface with no punctures, as it turns out to be rather intuitive. The inclusion of maximal and minimal punctures will be discussed in detail in the main part of the note.
Throughout this note we will focus on the range of non negative $U(1)_t$ symmetry flux (and zero flux only for $g>0$). In this range the above limit is the Coulomb limit in the language of ${\cal N}=2$, defined in terms of fugacities by setting $p,q,t\rightarrow 0$ while keeping $\frac{pq}{t}\equiv T$ fixed. $p$ and $q$ are the fugacities from the conventional definition of the superconformal index for ${\cal N}=1$ in 4d, see \cite{Rastelli:2016tbz} for details.\footnote{There is also a closely related limit of sending $p$ and $q$ to vanish while keeping $t$ fixed. This is the so called Hall-Littlewood limit focusing on the Higgs branch of the ${\cal N}=2$ theory. Allowing  fluxes in class ${\cal S}$ \cite{Bah:2012dg} leading to ${\cal N}=1$ theories, the two limits are closely related \cite{Beem:2012yn}. For different ranges of the fluxes at least one of these limits gives the freely generated ring we are interested in.} The fugacity $t$ is related to the $U(1)_t$ symmetry which in the case of $k=1$ is related to R symmetry of the enhanced ${\cal N}=2$ supersymmetry. 
For a general Riemann surface of genus $g$, general fluxes $b_i$, $c_j$ and $e$ for the symmetries $U(1)^{k-1}_{\beta_i}$, $U(1)^{k-1}_{\gamma_j}$, which are the Cartans of $SU(k)^2$, and $U(1)_t$, respectively, we find the following generating functions,

\be
\label{E:introFormula}
\mathcal{I}_{g,\left(b_{i},c_{j},e\right)}^{N,k} & = & PE\left[\sum_{i,j=1}^{k}\left(b_{i}-c_{j}+Ne+\left(N-1\right)\left(g-1\right)\right)\beta_{i}^{-N}\gamma_{j}^{N}T^{N}\right]\times\nonumber\\
 & & PE\left[\sum_{\ell=1}^{N-1}\left(\ell ke+\left(\ell k-1\right)\left(g-1\right)\right)T^{\ell k}\right],
\ee
where $PE$ is the plethystic exponent, see appendix \ref{A:pleth} for details. This implies that we have operators with charge $N$ under $U(1)_{\gamma _j}$, and $-N$ under $U(1)_t$ and $U(1)_{\beta_i}$, of multiplicity $b_i-c_j+Ne+(N-1)(g-1)$. These operators are bosonic if this number is positive and fermionic otherwise. In addition there are bosonic operators for $\ell=1,...,N-1$ numerated by $\ell k e+(\ell k-1)(g-1)$ with charge $-k\ell$ under $U(1)_t$. There is a choice of  R charge, not necessarily the conformal one, setting it for all of the above operators to be minus the $U(1)_t$ charge. One can provide an explanation of the formula by studying the flow of local operators in a certain freely generated sub ring of the six dimensional Higgs branch to four dimensions, which will be studied in \cite{babuip}.

Let us comment that the limit of the index we consider does not a priori correspond to an enhancement of supersymmetry. This is to be contrasted with the ${\cal N}=2$ case where the Coulomb limit of the index obtains contributions from states annihilated by more than one supercharge \cite{Gadde:2011uv}. The reason for this is that all three fugacities, $p,q,t$, in the ${\cal N}=2$ case correspond to charges on the right side of the anticommutator of supercharges. In the ${\cal N}=1$ case this is not true and moreover $t$ couples to flavor symmetry.  The states we count then are  a subset of the  states contributing in general to the index that are annihilated by one supercharge \cite{Romelsberger:2005eg, Kinney:2005ej,Dolan:2008qi}. Similar limits were considered in \cite{Spiridonov:2009za, 
Gaiotto:2015usa,Cremonesi:2015dja,
Hanany:2015via}.

This paper is organized as follows. In section \ref{S:quiverth} we list the various building blocks of theories in class ${\cal S}_k$ and methods to create new building blocks and combine them. From such building blocks we can deduce certain sphere and torus compactifications which have Lagrangian descriptions, and also properties of strongly coupled models by studying flows.
In section \ref{S:N=k=2 Coulomb} we give the initial hints from the Coulomb limit of the superconformal index  in theories with $N=k=2$ that lead to the conjectured formula. In section \ref{S:General N k} we give an ansatz for a formula of the  Coulomb limit of the index  for a Riemann surface with fluxes and no punctures, and generalize it in a self consistent way to general Riemann surfaces with fluxes, and maximal and minimal punctures. In section \ref{S:Constraints} we use known and new results for theories with $N=1,2,3$ and $k=1,2,3$, and completely determine the unknown functions in the ansatz given in section \ref{S:General N k}. Section \ref{S:Conj and Apps} summarizes the results.  Several appendices complement this paper. In particular we will give in complete detail Argyres-Seiberg type of dualities following from \cite{Gaiotto:2015usa} for two cases, $k=2$ and $N=3$, and $k=3$ and $N=2$. 

\section{4d ${\cal N}=1$ quiver theories}
\label{S:quiverth}

We consider quiver theories of class ${\cal S}_k$ as introduced in \cite{Gaiotto:2015usa}. We refer to the original paper for details and here we review the essential results.

The simple examples of theories in class ${\cal S}_k$ correspond to certain compactifications on a sphere with two maximal and any number of minimal punctures. These models can be built by gluing together theories corresponding to sphere with two maximal and one minimal punctures, which happen to be a collection of free fields, and thus are called a free trinion.

\begin{figure}[t]
        \centering
        \includegraphics[scale=0.26]{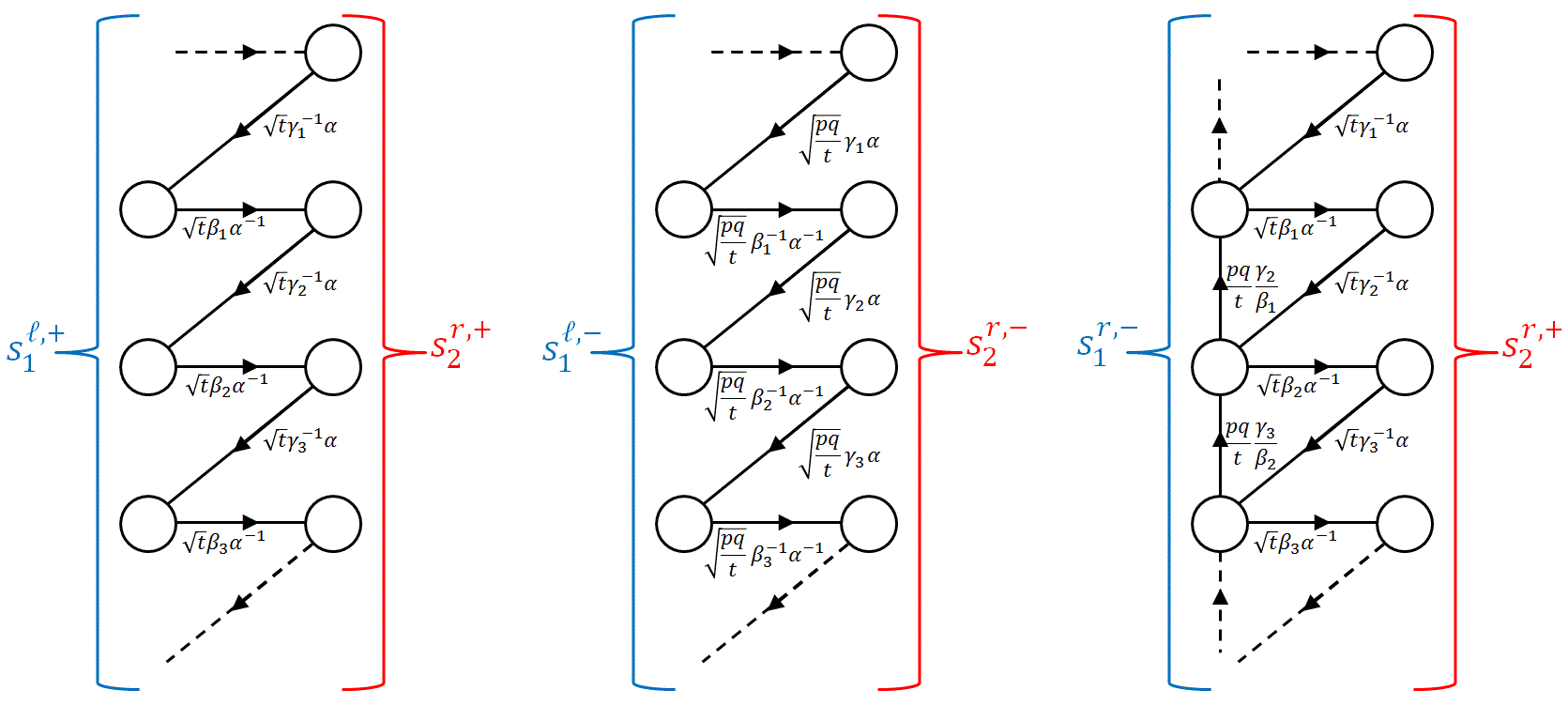}
        \caption{On the left we have a quiver diagram of a free trinion with all puncture signs positive. In the middle a free trinion with all punctures of negative sign. On the right a free trinion with left maximal puncture sign flipped to negative, this causes the orientation to flip as well. Each free trinion  would have two copies of $SU(N)^k$ flavor symmetry. The circles  represent $SU(N)$ groups, and there are $k$ of them winding around a circle on each side of the diagram, with arrows between them denoting bifundamental fields of these groups. $\alpha$ is the fugacity of a $U(1)$ symmetry related to the minimal puncture. In each diagram we indicate the maximal punctures attributes by $s_c^{o,\sigma}$ where $c$, $o$ and $\sigma$ are the color number, orientation and sign, respectively, in our notation. }
        \label{F:FT NFT FlippedP}
\end{figure}

More precisely, a free trinion theory corresponds  to a sphere with two maximal punctures, one minimal puncture, and flux $N_e=\frac{1}{2k}$ for $U(1)_t$ (this normalization is different than the one used in \cite{Gaiotto:2015usa,Razamat:2016dpl,Bah:2017gph}). The maximal and minimal punctures are related to the $SU(N)^k$ and $U(1)$ flavor symmetries, respectively. The maximal punctures are labeled by a color $c = 1,...,k$ a sign $\sigma=\pm 1$ and an orientation $o=\ell$ or $r$. We denote the two maximal punctures of the free trinion as left and right as they appear in the quiver diagram (Fig \ref{F:FT NFT FlippedP} left), this matches their orientation $o=\ell$ and $o=r$, respectively. These two maximal punctures have the same sign and a color shift of one, meaning if the color of the left puncture is $c=c_\ell$ then the color of the right puncture is $c=(c_\ell \mod \IZ_k)+1$. We assign $R$-charge $0$ to all the fields of the free trinion (this choice preserves some qualities of the Coulomb branch from ${\cal N}=2$, allowing for a simple form of its index).

One may want to examine a free trinion with punctures of negative sign (opposite to the one described above). This requires the following transformation of symmetries given using their fugacities
\be
\beta_i \rightarrow \beta_i^{-1} \ , \qquad \gamma_j \rightarrow \gamma_j^{-1} \ , \qquad t \rightarrow \frac{pq}{t},
\ee
where again $p$ and $q$ are the fugacities from the index definition of 4d ${\cal N}=1$ theories explained in \cite{Rastelli:2016tbz}. A quiver diagram of a negative sign free trinion is shown on figure \ref{F:FT NFT FlippedP} in the middle. One can also relate opposite signed maximal punctures to one another by adding bifundamental chiral multiplets as shown on figure \ref{F:FT NFT FlippedP} on the right. Once we use this procedure to flip the sign of a maximal puncture we get that its orientation is also changed.

\begin{figure}[t]
        \centering
        \includegraphics[scale=0.26]{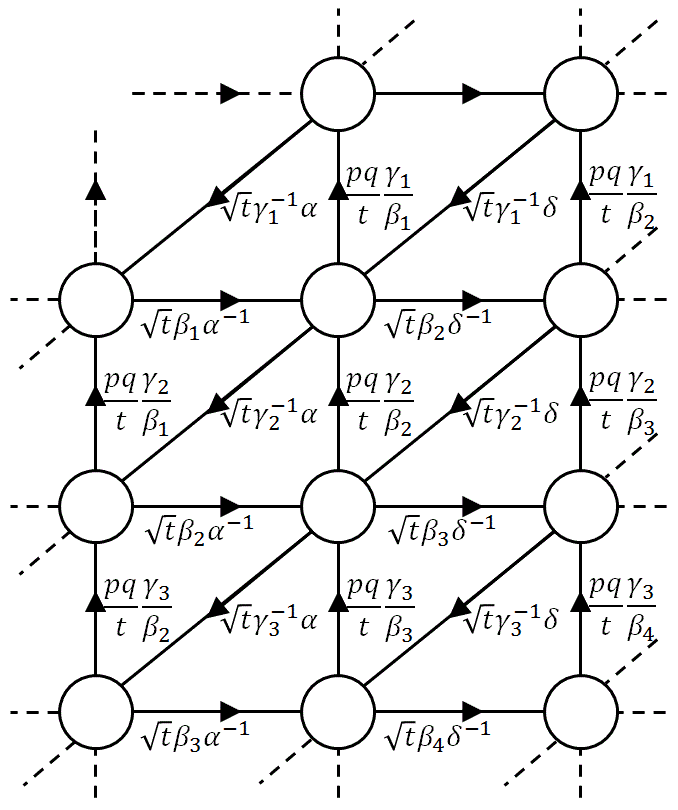} \hspace{2cm}
         \includegraphics[scale=0.23]{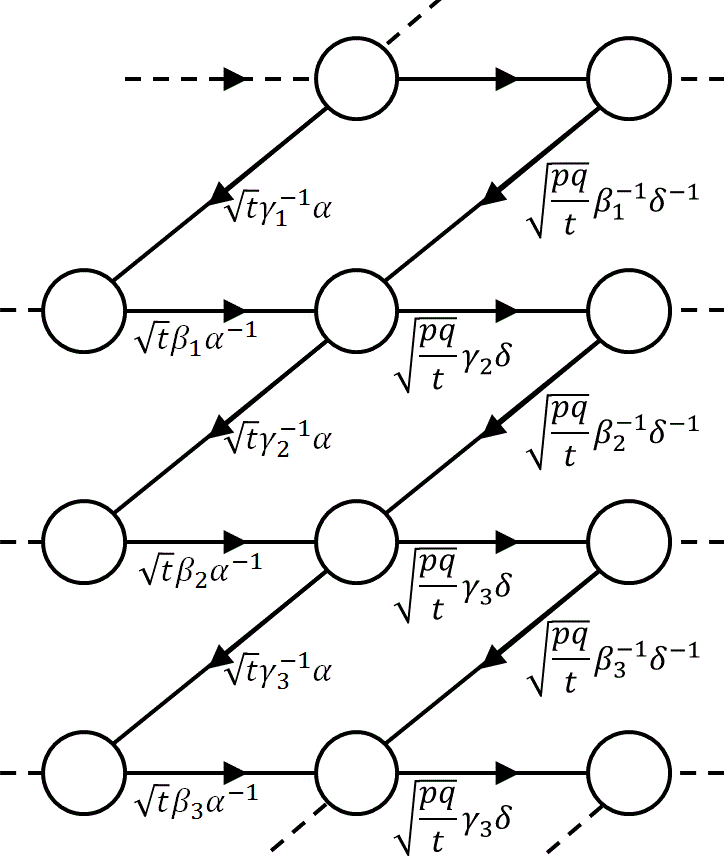}
        \caption{Quiver diagrams of the two types of gluings. On the left a $\Phi$-gluing along maximal punctures of color $c=2$ in our notation (the middle gluing). One can see the additional bifundamental chiral multiplets added in the gluing, denoted by upward arrows in the diagram. On the right an $S$-gluing of the same color $c=2$. }
        \label{F:Gluings}
\end{figure}

The next step one can use to find new theories, is to glue the different trinions described above, free trinions and flipped puncture free trinion. Such theories will obviously have more then three punctures. In general, gluing is associated with gauging the flavor symmetry of one maximal puncture at each trinion, as well as adding fields charged under the gauge symmetry which also  couple with a  superpotential to some operators in the glued theories. The gluing procedure is divided to two different versions depending on the signs of the maximal punctures we wish to glue:
\begin{itemize}
\item $\Phi$-gluing - This gluing is used when the two maximal punctures we wish to glue have the same sign (as well as same color and different orientation to preserve all internal symmetries). The gluing procedure requires adding an ${\cal N}=1$ $SU(N)^k$ vector multiplet, as well as bifundamental chiral multiplets between each two $SU(N)$ gauge groups along the gluing in a cyclic manner. These bifundamentals need to be coupled through the superpotential to the mesonic operators associated to the maximal punctures glued. This type of gluing is shown in a quiver diagram on the left in figure \ref{F:Gluings}.
\item $S$-gluing - This gluing is used  when the glued punctures have opposite sign (as well as same color and same orientation to preserve all internal symmetries). The gluing requires only adding an ${\cal N}=1$ $SU(N)^k$ vector multiplet, and coupling mesonic operators associated to both maximal punctures through  superpotential (Fig \ref{F:Gluings} on the right).
\end{itemize}

Another method of obtaining new theories, is closing minimal punctures. This is done by giving a vev to a baryon charged under the minimal puncture symmetry. The theory obtained after the flow corresponds to the same surface but with the puncture removed and the flux shifted.
The choice of the baryon to which we give a vev determines  the change in flux. Different choices lead in general to different   
fixed points. In addition gauge-singlet chiral multiplets need to be added, flipping the other baryons with same charge under the minimal puncture symmetry that didn't get a vev.

The last deformation we will use to find new theories is to close a maximal puncture. In general we can  close  maximal punctures  by giving vacuum expectation values to operators charged under the $SU(N)^k$ symmetry which corresponds to the puncture. Such a vev will break the symmetry to some subgroup depending on the choice of operators receiving the expectation value. 
 A very interesting case is to almost close the maximal puncture completely, leading to just a $U(1)$ symmetry, corresponding to  a minimal puncture. We will only consider the latter and refer the reader again to \citep{Gaiotto:2015usa} for a full description of the required vevs.

\section{$N=k=2$ Coulomb limit}
\label{S:N=k=2 Coulomb}

The $N=k=2$ case gives the simplest non-trivial theories of class ${\cal S}_k$. These theories and their superconformal indices have been studied thoroughly in \cite{Gaiotto:2015usa,Razamat:2016dpl,Bah:2017gph}. In addition to the quiver theories discussed in the previous section, here we also know what are the theories corresponding to spheres with three maximal punctures \cite{Razamat:2016dpl}. Knowing such theories we can construct models corresponding to any choice of flux and genus.
  The index of all these models can be computed as was done in \cite{Razamat:2016dpl} for many examples. 

Using these results, in the conventions defined in section \ref{S:quiverth}, one finds that the index has a simple Coulomb limit. The limit is taken by setting $p,q,t\goto 0$ while keeping $T\equiv \frac{pq}{t}$ finite, this leads to a trivial contribution, $1$,  of the chiral multiplets of the free trinion, and a simple non-trivial contribution of the chiral multiplets coming from the $\Phi$-gluing. The Coulomb limit of various theories is given in table \ref{T:N=k=2 indices}. Combining together theories in this table we can construct models of any genus and large variety of fluxes such that the $U(1)_t$ flux is positive.
When we consider the limit for a positive $U(1)_t$ flux model, the index is simply a product of the indices of the different models glued times the contributions of the $\Phi$ gluing.  
Note that although $V_i$ are obtained by gluing two free trinions, the only contribution they pick comes from the $\Phi$ gluing, since the index of the free trinons in our limit is one.  The $U(1)_\beta$ and $U(1)_\gamma$ flux tubes were computed in \cite{Bah:2017gph}, and as they have $U(1)_t$ flux zero the Coulomb limit of these is more complicated, however when they are glued to a theory with positive $U(1)_t$ flux they contribute simple multiplicative factor. That is the factor denoted in table \ref{T:N=k=2 indices} and it is not to be understood as the limit of the index of the tube models by themselves.

\begin{table}[t]
	\begin{center}
		\begin{tabular}{ | c || c | c | }
  		\hline			
  		Theory & ${\cal I}^C_{g,m,(s_1^{\ell,+},s_1^{r,+}),(s_2^{\ell,+},s_2^{r,+}),(b,c,e)}$ & Coulomb limit\\ \hline
		\hline
  		free trinion & ${\cal I}_{0,1,(1,0),(0,1),\left(0,0,\frac1{4}\right)}$&$1$ \qquad \\ \hline
		free trinion & ${ 		\cal I}_{0,1,(0,1),(1,0),\left(0,0,\frac1{4}\right)}$ & $1$ \\ \hline
  		$V_1$ & ${\cal I}_{0,2,(1,1),(0,0),\left(0,0,\half\right)}
		$

		&
		$\frac1{(1-T^2)\left(1-T^2 (\beta^{-1}\gamma)^{\pm 2}\right)}$ \\ \hline
  		$V_2$ & ${\cal I}_{0,2,(0,0),(1,1),\left(0,0,\half\right)}$                    &                  $\frac1{(1-T^2)\left(1-T^2 (\beta\gamma)^{\pm 2}\right)}$ \\ \hline
  		$T_A$ & ${\cal I}_{0,0,(0,0),(2,1),\left(-\frac1{4},-\frac1{4},\half\right)}$                    &                  $\frac1{\left(1-T^2 (\beta\gamma)^{-2}\right)\left(1-T^2 (\beta\gamma)^{\pm 2}\right)}$ \\ \hline
  		$T_B$ & ${\cal I}_{0,0,(1,0),(1,1),\left(\frac1{4},-\frac1{4},\half\right)}$                    &                  $\frac1{\left(1-T^2(\beta^{-1}\gamma)^{-2}\right)\left(1-T^2(\beta\gamma)^{\pm 2}\right)}$ \\ \hline
  		$U(1)_\beta$ flux tube & ${\cal I}_{0,0,(0,0),(1,1),\left(-1,0,0\right)}$                    &                  $\frac{\left(1-T^{2}\right)\left(1-T^{2}\left(\beta^{-1}\gamma\right)^{-2}\right)^{2}\left(1-T^{2}\left(\beta\gamma\right)^{2}\right)}{\left(1-T^{2}\left(\beta\gamma\right)^{-2}\right)}$ \\ \hline
  		$U(1)_\gamma$ flux tube & ${\cal I}_{0,0,(0,0),(1,1),\left(0,-1,0\right)}$                    &                  $\frac{\left(1-T^{2}\right)\left(1-T^{2}\left(\beta^{-1}\gamma\right)^{2}\right)^{2}\left(1-T^{2}\left(\beta\gamma\right)^{2}\right)}{\left(1-T^{2}\left(\beta\gamma\right)^{-2}\right)}$ \\ \hline
  		$U(1)_t$ flux tube & ${\cal I}_{0,0,(0,0),(1,1),\left(0,0,1\right)}$                    &                  $\frac1{(1-T^2)\left(1-T^2(\beta^{-1}\gamma)^{\pm 2}\right)\left(1-T^2(\beta\gamma)^{\pm 2}\right)^2}$ \\
  		\hline 
		\end{tabular}
	\end{center}
	\caption{The Coulomb index of various $N=k=2$ theories. Here $g$ is the genus of the Riemann surface, $m$ the number of minimal punctures, $s_c^{o,\sigma}$ the number of maximal punctures of color $c$ orientation $o$ and sign $\sigma$, and $b$, $c$ and $e$ are the $\beta$, $\gamma$, and $t$ associated fluxes, respectively.}
	\label{T:N=k=2 indices}
\end{table} 
 
Using the known results of the index in the Coulomb limit, one can combine them to find the index for a general Riemann surface with general flux. 
 The index for a general Riemann surface, number of punctures and fluxes is given for $N=k=2$ by
\be
\label{E:N=k=2 coulomb}
\mathcal{I}^{C}_{(g,m,s^+_{1},s^+_{2},\left(b_i,c_j,e\right)} & = & PE\left[\sum_{i,j=1}^{2}\left(\frac{1}{2}(\delta_{i=j}s^+_{1}+\delta_{i\ne j}s^+_{2})-b_{i}+c_{j}+2e+g-1\right)\beta_{i}^{-2}\gamma_{j}^{2}T^{2}\right]\times\nonumber\\
 & & PE\left[\left(\frac{1}{2}m+2e+g-1\right)T^{2}\right],
\ee
where $\beta_1=\beta_2^{-1}=\beta$, $\gamma_1=\gamma_2^{-1}=\gamma$, $s_c^+=s_c^{\ell,+}+s_c^{r,+}$, $b_i=(b,-b)$ and $c_i=(c,-c)$.

We stress again that for sufficiently low (negative or vanishing) $U(1)_t$ flux the limit of the index ceases to be a simple product of factors corresponding to the basic ingredients. It becomes analogous to the Hall-Littlewood limit of the index in the class ${\cal S}$ case \cite{Gadde:2011uv}.  However, if theories with negative flux and positive are combined so that the total flux is positive, the index again takes a very simple form given by the expression here. The  limit we are considering can give the above simple expression even when $U(1)_t$ flux is zero or negative if the genus is high enough. However, we will state the conjecture here only for $e>0$ as this is the range of fluxes that we have thoroughly studied. 

One can in principle perform similar computations for other compactifications in class ${\cal S}_k$ though the theories corresponding to higher genus and more punctures are not known. The way one can proceed is to study the procedure of closing punctures and conjecture dualities relating theories we reviewed in the previous section to theories involving new models. However, this procedure is rather complicated. We derive in Appendix B two examples of such dualities and we will use those in the following sections. We will next conjecture a generalization of the Coulomb limit for higher $N$ and $k$ and test it against all the known results. An important guide-line for the generalization will be to look for an expression which is a plethystic exponent of generators whose multiplicity is linearly dependent on flux and all the parameters. 
This assumption is motivated by the explicit computations and by the six dimensional logic of \cite{babuip} (see \cite{Beem:2012yn} and \cite{Kim:2017toz} Appendix E for a summary of the idea behind this logic).

\section{Coulomb limit for general N and k}
\label{S:General N k}

\begin{figure}[t]
        \centering
        \includegraphics[scale=0.2]{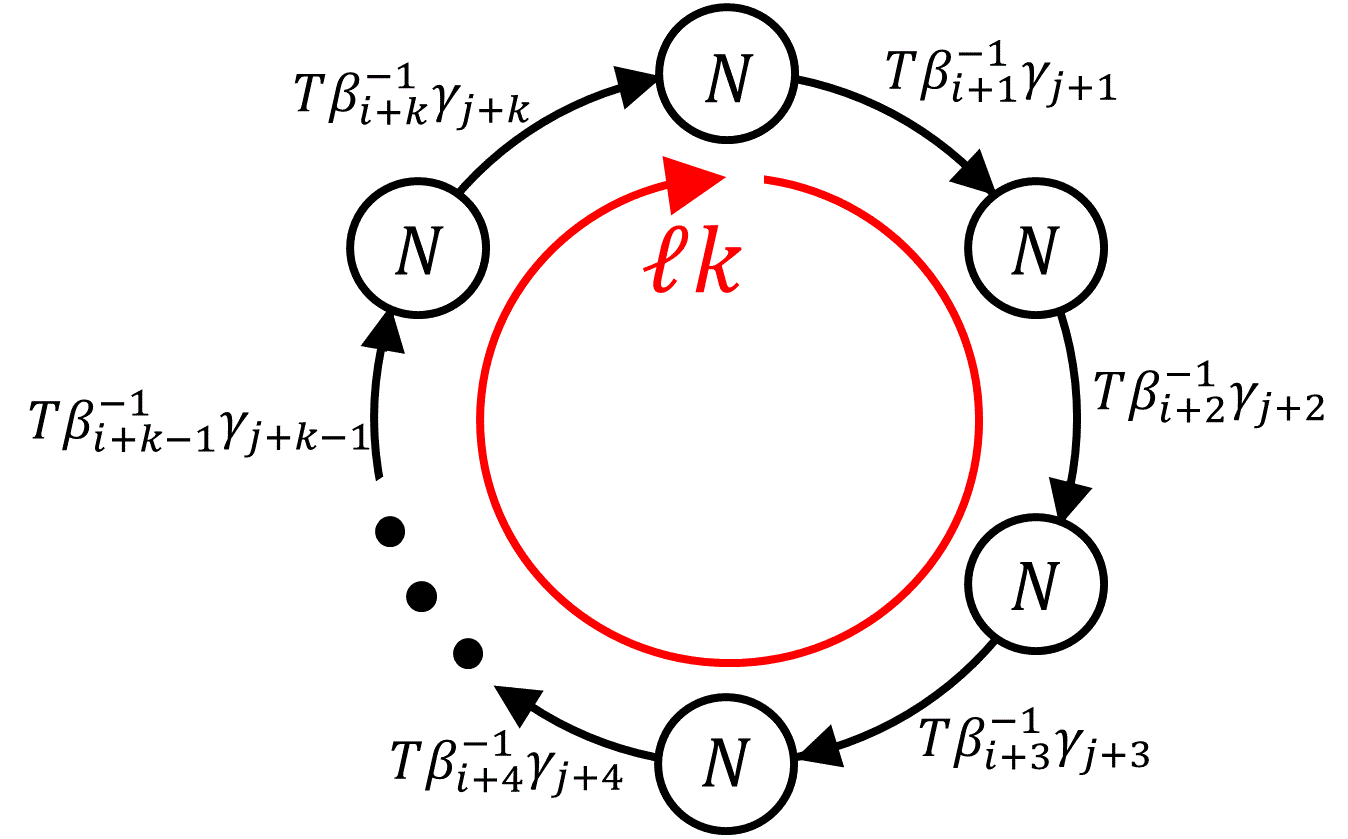}
        \caption{When we glue surfaces by gauging puncture symmetry with $\Phi$ gluing the gauge groups form a circle. 
 The operators we see in our formula can be naturally seen already in this gluing.   One can build two kinds of gauge invariant objects out of this  ring. Baryons such as $T^N \beta_{i+m}^{-N} \gamma_{j+m}^N$, and integral multiples of cycles up to $N-1$ cycles as indicated by the red arrow, with contribution $T^{\ell k}$, where $T^k$ is the contribution from one cycle and $\ell=1,..,N-1$ the number of cycles, remembering that $\prod_{i=1}^k \beta_i=\prod_{j=1}^k \gamma_j =1$. Different colors will give rise to different types of operators charged under the $\beta$ and $\gamma$ symmetries. 
   }
        \label{F:Cring}
\end{figure}

Equation \eqref{E:N=k=2 coulomb} is suggestive of a general form of the index in the Coulomb limit. For a Riemann surface with fluxes and no punctures (for simplicity) we suggest the following  ansatz for the Coulomb limit with general $N$ and $k$

\be
\label{E:Ansatz}
\mathcal{I}_{g,\left(b_{i},c_{j},e\right)}^{N,k} & = & PE\left[\sum_{i,j=1}^{k}\left(f_{b}\left(N,k\right)b_{i}+f_{c}\left(N,k\right)c_{j}+f_{e}\left(N,k\right)e+f_{g}\left(N,k\right)\left(g-1\right)\right)\beta_{i}^{-N}\gamma_{j}^{N}T^{N}\right]\nonumber\\
 & & \times PE\left[\sum_{\ell=1}^{N-1}\left(h_{e}\left(\ell ,k\right)e+h_{g}\left(\ell ,k\right)\left(g-1\right)\right)T^{\ell k}\right],
\ee
where the functions $f_i(N,k)$ and $h_j(k,\ell)$ with $i=b,c,e,g$ and $j=e,g$ need to be determined using theories with various $N$ and $k$. The powers of $\beta_i$, $\gamma_j$ and $T$ are motivated 
by the generalization of the operators appearing in gluing as is illustrated
 in Figure \ref{F:Cring}. The charges of operators that appear in the limit are those of operators one can see in the gluing of different punctures. In particular we have operators winding around the affine circle of $A_{k-1}$ with charges being multiples of $k$, and
 baryonic operators which have charges scaling with $N$.  The details of the charges depend on the colors and signs of punctures. The conjecture is that only such types of operators appear in the Coulomb limit with the multiplicities determined by the fluxes, types of punctures, and type of Riemann surface. We need to determine then these multiplicities as function of fluxes, genus, and other parameters.

The information we can use to fill in the blanks is limited. Two pieces of information we can immediately insert to this formula for any $k$ and $N$ is the contribution of the free trinion and the contribution of the $\Phi$-gluing. These contributions won't help us find the unknown functions, but will allow us to add maximal and minimal punctures to the ansatz. The remaining information will come from our knowledge on how to close minimal and maximal punctures, and the knowledge on how to flip the sign of a puncture, as disscused in section \ref{S:quiverth} and prescribed by \cite{Gaiotto:2015usa}. Thus, in order to make it simpler to find the unknown functions we will add punctures to \eqref{E:Ansatz}.

\subsection*{Adding maximal punctures}

First, the contribution of the $\Phi$-gluing is required in order to add maximal punctures. This is given for general $N$, $k$ and color by
\be
V_{c}^{N,k}=PE\left[\sum_{i=1}^{k}\beta_{i+c-1}^{-N}\gamma_{i+1}^{N}T^{N}+\sum_{\ell=1}^{N-1}T^{\ell k}\right],
\ee
where $\beta_i$ and $\gamma_j$ are defined cyclically, meaning $\beta_i=\beta_{i+k}$ and $\gamma_j=\gamma_{j+k}$. 
Now, we can remove one handle from a general Riemann surface, and create two maximal punctures of color $c$ and opposite orientation $\ell$ and $r$. This can be executed $s_c$ times for each color $c$, the result is
\be
\label{E:Ansatz w punctures}
\mathcal{I}_{g,s_{c}^{\ell,+}=s_{c}^{r,+}=s_{c},\left(b_{i},c_{j},e\right)}^{N,k} & = & PE\left[\sum_{i,j=1}^{k}\left(f_{b}\left(N,k\right)b_{i}+f_{c}\left(N,k\right)c_{j}+f_{e}\left(N,k\right)e\right)\beta_{i}^{-N}\gamma_{j}^{N}T^{N}\right]\times\nonumber\\
 & & PE\left[\sum_{i,j=1}^{k}\left(f_{g}\left(N,k\right)\left(g-1+\frac{s_{tot}}{2}\right)-\sum_{c=1}^{k}s_{c}\delta^k_{i-c+2,j}\right)\beta_{i}^{-N}\gamma_{j}^{N}T^{N}\right]\times\nonumber\\
 & & PE\left[\sum_{\ell=1}^{N-1}\left(h_{e}\left(\ell ,k\right)e+h_{g}\left(\ell ,k\right)\left(g-1+\frac{s_{tot}}{2}\right)-\frac{s_{tot}}{2}\right)T^{\ell k}\right],
\ee
where $s_{tot}=s_c^{\ell,+}+s_c^{r,+}$, and we defined $\delta^k_{i,j}=\delta_{(i\mod k),(j\mod k)}$. A number of these theories can be combined to find the same result as in \eqref{E:Ansatz w punctures}, showing self consistency.

Next, we want to find what is the formula for unequal number of left and right orientation maximal punctures. This means we need to divide the contribution of two opposite orientation maximal punctures to two parts. It is not clear how this should be done, and if it should involve some non-trivial opposite sign contributions that would cancel when we add them up. This can only be determined by self consistency conditions. We find the self consistent result to be
\be
\label{E:Ansatz w o punctures}
\mathcal{I}_{g,s_{c}^{r},s_{c}^{\ell},\left(b_{i},c_{j},e\right)}^{N,k} & = & PE\left[\sum_{i,j=1}^{k}\left(f_{b}\left(N,k\right)b_{i}+f_{c}\left(N,k\right)c_{j}+f_{e}\left(N,k\right)e\right)\beta_{i}^{-N}\gamma_{j}^{N}T^{N}\right]\times\nonumber\\
 & & PE\left[\sum_{i,j=1}^{k}\left(f_{g}\left(N,k\right)\left(g-1+\frac{s_{tot}}{2}\right)\right)\beta_{i}^{-N}\gamma_{j}^{N}T^{N}\right]\times\nonumber\\
 & & PE\left[\sum_{i,j,c,n=1}^{k}\frac{k-2n}{2k}\left(s_{c}^{r}\delta_{i+2-c+n,j}+s_{c}^{\ell}\delta_{i+2-c-n,j}\right)\beta_{i}^{-N}\gamma_{j}^{N}T^{N}\right]\times\nonumber\\
 & & PE\left[\sum_{\ell=1}^{N-1}\left(h_{e}\left(\ell ,k\right)e+h_{g}\left(\ell ,k\right)\left(g-1+\frac{s_{tot}}{2}\right)-\frac{s_{tot}}{2}\right)T^{\ell k}\right],
\ee
where we dropped the $+$ sign from $s_{c}^{\ell,+}$ and $s_{c}^{r,+}$ to abbreviate the notation, it should be clear from now on that we refer to positive sign punctures unless we state otherwise explicitly.  
We will show that this formula is self consistent after we add minimal punctures.

\subsection*{Adding minimal punctures}

Finally, we want to add minimal punctures as well. In order to add minimal punctures we need to know the contribution in the Coulomb limit of theories with two maximal punctures with opposite orientation of color $c$ and $k$ minimal punctures, which is given by
\be
\mathcal{I}_{,s_{c}^{r}=s_{c}^{\ell}=1,m=k,\left(0,0,\frac{k}{2k}\right)}^{N,k}=PE\left[\sum_{i,j=1}^{k}\left(1-\delta_{i-c+2,j}\right)\beta_{i}^{-N}\gamma_{j}^{N}T^{N}+\sum_{\ell=1}^{N-1}\left(k-1\right)T^{\ell k}\right]
\ee
where $\frac{1}{2k}$ is the $e$ flux of a single free trinion. Adding $m$ minimal punctures by adding $m/k$ such theories glued with $V_{c}^{N,k}$ contributions, we find
\be
\label{E:AnsatzPunc}
\mathcal{I}_{g,s_{c}^{r},s_{c}^{\ell},m,\left(b_{i},c_{j},e\right)}^{N,k} & = & PE\left[\sum_{i,j=1}^{k}\left(f_{b}\left(N,k\right)b_{i}+f_{c}\left(N,k\right)c_{j}+f_{e}\left(N,k\right)\left(e-\frac{m}{2k}\right)\right)\beta_{i}^{-N}\gamma_{j}^{N}T^{N}\right]\times\nonumber\\
 & & PE\left[\sum_{i,j=1}^{k}\left(f_{g}\left(N,k\right)\left(g-1+\frac{s_{tot}}{2}\right)+\frac{m}{k}\right)\beta_{i}^{-N}\gamma_{j}^{N}T^{N}\right]\times\nonumber\\
 & & PE\left[\sum_{i,j=1}^{k}\left(\sum_{c=1}^{k}\sum_{n=1}^{k}\frac{k-2n}{2k}\left(s_{c}^{r}\delta_{i+2-c+n,j}+s_{c}^{\ell}\delta_{i+2-c-n,j}\right)\right)\beta_{i}^{-N}\gamma_{j}^{N}T^{N}\right]\times\nonumber\\
 & & PE\left[\sum_{\ell=1}^{N-1}\left(h_{e}\left(\ell ,k\right)\left(e-\frac{m}{2k}\right)+h_{g}\left(\ell ,k\right)\left(g-1+\frac{s_{tot}}{2}\right)\right)T^{\ell k}\right]\times\nonumber\\
 & & PE\left[\sum_{\ell=1}^{N-1}\left(m-\frac{s_{tot}}{2}\right)T^{\ell k}\right]
\ee

\subsection*{Self consistency check}

In order to make sure the formula we found is self consistent, we use a non-trivial test case. We expect the gluing contribution of color $c=1$ to be given by a theory with two maximal punctures of adjacent colors to $c=1$, meaning $c=2$ right orientation puncture and $c=k$ left orientation puncture. In addition to two minimal punctures, and a non-trivial flux of $\frac1{k}$ only for $e$. An explicit calculation shows that the Coulomb index reduces to
\be
\mathcal{I}_{g=0,s_{2}^{r}=s_{k}^{\ell}=1,m=2,\left(0,0,\frac1{k}\right)}^{N,k} = PE\left[\sum_{i=1}^{k}\beta_{i}^{-N}\gamma_{i+1}^{N}T^{N}+\sum_{\ell=1}^{N-1}T^{\ell k}\right],
\ee
which is exactly the contribution of the gluing as we expected. This seemingly trivial check fails to the best of our knowledge for any choice of left and right orientation contribution except the one chosen in \eqref{E:Ansatz w o punctures}.

\section{Ansatz constraints}
\label{S:Constraints}
As we stated earlier we want to use the superconformal index Coulomb limit of theories with $N=1,2,3$ and $k=1,2,3$ to impose constraints on the ansatz \eqref{E:Ansatz} and \eqref{E:AnsatzPunc}, allowing us to find the unknown functions in it. For $k=1$ and $N=k=2$ we will compare \eqref{E:Ansatz} to known similar formulas. For the other two cases $N=2,k=3$ and $N=3,k=2$ we will find the Coulomb limit of various theories using our knowledge on how to close and flip punctures. We will find the indices of $\beta$ and $\gamma$ tubes/tori using our knowledge on how to close minimal punctures. Also we'll find the index of joined free trinions with one maximal puncture closed using our knowledge on how to close maximal punctures. At last we'll calculate the index of a free trinion with one puncture flipped and completely closed. That will add the last bit of information needed to completely determine the unknown functions in the ansatz.
\subsection*{$k=1$ (class $\mathcal{S}$) constraints}
Starting with the simplest case of $k=1$ also known as class ${\cal S}$ we quote the result of \cite{Beem:2012yn} written in our notation, giving the general form of the Coulomb limit for any $N$
\be
\label{E:Coulomb k=1}
\mathcal{I}_{g,\left(e\right)}^{N} = PE\left[\left(Ne+\left(N-1\right)\left(g-1\right)\right)T^{N}+\sum_{\ell=2}^{N-1}\left(\ell e+\left(\ell-1\right)\left(g-1\right)\right)T^{\ell}
\right].
\ee
Comparing \eqref{E:Coulomb k=1} with \eqref{E:Ansatz} for $k=1$ we find the following constraints 
\be
\label{E:constraints k=1}
f_{e}\left(N,k=1\right)=N\ ,\, f_{g}\left(N,k=1\right)=N-1\ ,\, 
h_{e}\left(\ell ,k=1\right)=\ell\ ,\, h_{g}\left(\ell ,k=1\right)=\ell-1.
\ee
One should note that the $\ell=1$ case is missing from \eqref{E:Coulomb k=1}. This is due to the simple fact that the $\ell=1$ contribution when reduced to the $k=1$ case arises from a free chiral multiplet, which is removed in class ${\cal S}$.
\subsection*{$N=k=2$ constraints}
Comparing \eqref{E:N=k=2 coulomb} without punctures with \eqref{E:Ansatz} for $k=N=2$ we find the constraints to be
\be
\label{E:constraints k=N=2}
-f_{b}\left(2,2\right)=f_{c}\left(2,2\right)=f_{g}\left(2,2\right)=h_{g}\left(1,2\right)=1\quad;\quad f_{e}\left(2,2\right)=h_{e}\left(1,2\right)=2
\ee
where the first argument of $f_i$ and $h_j$ is $N$ and $\ell$, respectively, and the second argument is $k$.
\subsection*{$N=2$ $k=3$ constraints}
\subsubsection*{Flux tubes/tori}
Computing the index for $\beta$ and $\gamma$ flux tori or flux tubes glued to a free trinion, with non trivial fluxes $b=(2,-1,-1)$ and $c=(-2,1,1)$, respectively, we get the following Coulomb limit
\be
{\cal I}_{\beta_1 tori}^C & = & \mathcal{I}_{g=1,\left(b_{i}=(2,-1,-1),c_{j}=0,e=0\right)}^{N=2,k=3} = \frac{\prod_{i=1}^{3}\left(1-T^{2}\left(\beta_{1}^{-1}\gamma_{i}\right)^{2}\right)^{2}}{\prod_{i=1}^{3}\left(1-T^{2}\left(\beta_{2}^{-1}\gamma_{i}\right)^{2}\right)\left(1-T^{2}\left(\beta_{1}\beta_{2}\gamma_{i}\right)^{2}\right)} \nonumber\\
{\cal I}_{\gamma_1 tori}^C & = & \mathcal{I}_{g=1,\left(b_{i}=0,c_{j}=(-2,1,1),e=0\right)}^{N=2,k=3} = \frac{\prod_{i=1}^{3}\left(1-T^{2}\left(\beta_{i}^{-1}\gamma_{1}\right)^{2}\right)^{2}}{\prod_{i=1}^{3}\left(1-T^{2}\left(\beta_{i}^{-1}\gamma_{2}\right)^{2}\right)\left(1-T^{2}\left(\beta_{i}^{-1}\gamma_{3}\right)^{2}\right)}.
\ee
Comparing with \eqref{E:Ansatz} we find that the $b_i$, $c_j$ and $e$ fluxes are constrained to be as chosen up to an overall factor which we determined. We can write the general flux change for closing a minimal puncture for $N=2$ by
\be
\label{E:CminP N=2}
CminP_{flux}=\begin{cases}
\left(\left(b_{i}=\pm\frac{k-1}{k},b_{j\ne i}=\mp\frac{1}{k}\right),\vec{0},-\frac{1}{2k}\right) & vev\propto\beta_{i}^{\pm 1}\\
\left(\vec{0},\left(c_{i}=\mp\frac{k-1}{k},c_{j\ne i}=\pm\frac{1}{k}\right),-\frac{1}{2k}\right) & vev\propto\gamma_{i}^{\pm 1}
\end{cases},
\ee 
where the signs are correlated. In the degenerate case of $k=1$ the contribution to the $e$ flux will be as stated ($-\frac12$) for positive minimal punctures regardless of the vev (there is only one possibility for a vev in this case). In general, negative sign minimal punctures will get an opposite sign contribution to the $e$ flux.
The comparison also leads to the following constraints
\be
\label{E:tube constraints k=3 N=2}
-f_{b}\left(2,3\right)=f_{c}\left(2,3\right)=1.
\ee
Again, assuming linear dependence on $k$, we can immediately deduce from \eqref{E:constraints k=N=2} and \eqref{E:tube constraints k=3 N=2}
\be
\label{E:fb and fc N=2}
-f_{b}\left(N=2,k\right)=f_{c}\left(N=2,k\right)=1.
\ee

\subsubsection*{Closing maximal puncture}
From gluing three free trinions together and closing one maximal puncture to a minimal one we obtain a "tail" that can be used to find the interacting trinion through S-duality, see appendix \ref{A:intTrin} for the duality in terms of the index, and anomaly matching to 6d. The Coulomb limit of the index of this theory is given by\footnote{In some cases such as this, the Coulomb limit is seemed to be not well defined at first, due to several infinities and zeroes. This can be mended by performing an appropriate Seiberg duality, leading to an index in which the Coulomb limit is manifestly well defined.}
\be
\label{E:tail k=3 N=2}
\mathcal{I}_{g=0,s_{1}^{r}=1,m=4,\left(b_{i},c_{j},e\right)}^{N=2,k=3} = \frac{\left(1-T^{2}\left(\gamma_{2}\beta_{1}^{-1}\right)^{2}\right)\left(1-T^{2}\left(\gamma_{3}\beta_{2}^{-1}\right)^{2}\right)\left(1-T^{2}\left(\gamma_{2}\beta_{2}^{-1}\right)^{2}\right)}{\left(1-T^{3}\right)},
\ee
where we didn't assume anything about the fluxes, since closing a maximal puncture by giving a vev to some mesons doesn't necessarily give the same flux for different $k$ and $N$. By comparing \eqref{E:tail k=3 N=2} to \eqref{E:AnsatzPunc} we can determine the fluxes to be $\left(\vec{b}=(0,\frac{1}{3},-\frac{1}{3}),\vec{c}=(\frac{1}{3},-\frac{1}{3},0),e\right)$, and we can express $e$ as the initial flux minus the shift caused by the closure of the maximal puncture $e=\frac{1}{2}-e_{maxC}\left(N=2,k=3\right)$. $e_{maxC}\left(N,k\right)$ should be an integer multiplication of $\frac{1}{2k}$, and also proportional to $k-1$ to ensure the closure of maximal punctures in class ${\cal S}$ does not shift the flux as known. Adding our knowledge on the t-flux shift when closing maximal punctures in models with $N=k=2$, we can set
\be
\label{E:Cmax shift N=2}
e_{maxC}\left(N=2,k\right)=\frac{k-1}{2k}.
\ee
Thus, setting $e=1/6$ for the above "tail". Using these fluxes and the known results for $N=k=2$ from table \ref{T:N=k=2 indices}, we can determine the flux change from closing a maximal puncture to a minimal puncture for $N=2$ and general $k$ given by
\be
\label{E:CmaxP N=2}
CmaxP_{flux}(N=2)=\begin{cases}
\left(\left(b_{i}=\pm\frac{k-1}{k^{2}},b_{j\ne i}=\mp\frac{1}{k^{2}}\right),\vec{0},0\right) & \mbox{each meson vev}\propto\beta_{i}^{\pm 1}\\
\left(\vec{0},\left(c_{i}=\pm\frac{k-1}{k^{2}},c_{j\ne i}=\mp\frac{1}{k^{2}}\right),0\right) & \mbox{each meson vev}\propto\gamma_{i}^{\pm 1}\\
\left(\vec{0},\vec{0},\mp \frac{1}{k^{2}}\right) & \mbox{each meson vev} \propto t^{\pm 1}
\end{cases}
\ee
where the signs are correlated. By each meson vev we include the initial vevs given in order to close the maximal puncture as well as the subsequent meson vevs in the next lines of trinions forced by the initial ones, for an extensive discussion on maximal puncture closing procedure see \cite{Gaiotto:2015usa}. For $N=2$ there are $k(k-1)/2$ such mesons giving a total shift in $e$ of $(k-1)/2k$ as expected. In order to clarify the flux shift process we'll use \eqref{E:tail k=3 N=2} as an example. We start from three free trinions glued together, with a flux of $\left(b=\vec{0},c=\vec{0},e=\half\right)$. The initial vevs where given to mesons $u_{1}u_{2}^{-1}t\beta_{1}\gamma_{2}^{-1}$ and $u_{2}u_{3}^{-1}t\beta_{2}\gamma_{3}^{-1}$, these caused the subsequent meson $z_{1}z_{2}^{-1}t\beta_{2}\gamma_{2}^{-1}$ on the next trinion to get a vev as well. The meson vevs change of flux is
\be
u_{1}u_{2}^{-1}t\beta_{1}\gamma_{2}^{-1} & \rightarrow & \left(b=(\frac{2}{9},-\frac1{9},-\frac1{9}),c=(\frac1{9},-\frac{2}{9},\frac1{9}),e=-\frac1{9}\right)\nonumber\\
u_{2}u_{3}^{-1}t\beta_{2}\gamma_{3}^{-1} & \rightarrow & \left(b=(-\frac1{9},\frac{2}{9},-\frac1{9}),c=(\frac1{9},\frac1{9},-\frac{2}{9}),e=-\frac1{9}\right)\nonumber\\
z_{1}z_{2}^{-1}t\beta_{2}\gamma_{2}^{-1} & \rightarrow & \left(b=(-\frac1{9},\frac{2}{9},-\frac1{9}),c=(\frac1{9},-\frac{2}{9},\frac1{9}),e=-\frac1{9}\right).
\ee
Summing these changes of flux along with the initial flux we recover the "tail" flux. Besides the flux constraints we also find
\be
\label{E:CmaxP constraints k=3 N=2}
f_{e}\left(2,3\right)+f_{g}\left(2,3\right) & = & 3\nonumber\\
h_{e}\left(1,3\right)+h_{g}\left(1,3\right) & = & 5.
\ee
These constraints are consistent with \eqref{E:constraints k=1} and \eqref{E:constraints k=N=2}.

\subsubsection*{Flipping a maximal puncture and closing it completely}
Using the knowledge on how to flip the sign of a maximal puncture from \cite{Gaiotto:2015usa}, we can flip one of the maximal punctures of the free trinion and close it completely. The Coulomb index of this theory is
\be
\mathcal{I}_{g=0,s_{2}^{r}=1,m=1,\left(b_i,c_j,e\right)}^{N=2,k=3} & = & \left(\left(1-\beta_{1}^{-2}\gamma_{1}^{2}T^{2}\right)\left(1-\beta_{1}^{-2}\gamma_{2}^{2}T^{2}\right)\left(1-\beta_{1}^{-2}\gamma_{3}^{2}T^{2}\right)\left(1-T^{3}\right)\right)^{-1}\nonumber\\
 & & \left(\left(1-\beta_{2}^{-2}\gamma_{3}^{2}T^{2}\right)\left(1-\beta_{2}^{-2}\gamma_{1}^{2}T^{2}\right)\left(1-\beta_{3}^{-2}\gamma_{1}^{2}T^{2}\right)\right)^{-1},
\ee
with flux
\be
\left(\vec{b}=\left(-\frac{1}{3},0,\frac{1}{3}\right),\vec{c}=\left(\frac{1}{3},-\frac{1}{3},0\right),e=\frac{2}{3}\right).
\ee
Comparing to \eqref{E:AnsatzPunc} leads to the following constraints
\be
\label{E:Flip and CmaxP constraints k=3 N=2}
f_{e}\left(2,3\right)-f_{g}\left(2,3\right) & = & 1\nonumber\\
h_{e}\left(1,3\right)-h_{g}\left(1,3\right) & = & 1,
\ee
where again these constraints are consistent with \eqref{E:constraints k=1} and \eqref{E:constraints k=N=2}.

\subsection*{$N=3$ $k=2$ constraints}

\subsubsection*{Flux tubes/tori}

$\beta$ and $\gamma$ flux tori with fluxes $b=(-1,1)$ and $c=(-1,1)$, respectively, have the following Coulomb indices
\be
\mathcal{I}_{g=1,\left(b_{i}=(-1,1),c_{j}=(0,0),e=0\right)}^{N=3,k=2}=\frac{\left(1-T^{3}\left(\beta^{-1}\gamma\right)^{-3}\right)\left(1-T^{3}\left(\beta\gamma\right)^{3}\right)}{\left(1-T^{3}\left(\beta^{-1}\gamma\right)^{3}\right)\left(1-T^{3}\left(\beta\gamma\right)^{-3}\right)}\nonumber\\
\mathcal{I}_{g=1,\left(b_{i}=(0,0),c_{j}=(-1,1),e=0\right)}^{N=3,k=2}=\frac{\left(1-T^{3}\left(\beta^{-1}\gamma\right)^{3}\right)\left(1-T^{3}\left(\beta\gamma\right)^{3}\right)}{\left(1-T^{3}\left(\beta^{-1}\gamma\right)^{-3}\right)\left(1-T^{3}\left(\beta\gamma\right)^{-3}\right)},
\ee
where we determined the flux assuming \eqref{E:CminP N=2} holds for all $N$. Comparing with \eqref{E:Ansatz}, we find the constraints
\be
\label{E:tori constraints N=3 k=2}
-f_{b}\left(3,2\right)=f_{c}\left(3,2\right)=1.
\ee
Thus we can come the conclusion that the change of flux caused by closing a minimal puncture is given by \eqref{E:CminP N=2}, and is independent of $N$ as expected. Also, considering \eqref{E:fb and fc N=2} and \eqref{E:tori constraints N=3 k=2} we can determine that
\be
\label{E:fb and fc}
-f_{b}\left(N,k\right)=f_{c}\left(N,k\right)=1.
\ee

\subsubsection*{Closing maximal puncture}
We obtain a "tail" from gluing four free trinions ($(N-1)k$ free trinions in general) and closing a maximal puncture. This "tail" can again be used to find the interacting trinion through S-duality, see appendix \ref{A:intTrin}. The Coulomb index is found to be\footnote{This computation is very involved, and requires to preform three Seiberg dualities to get a manifestly well defined Coulomb limit. After that it is still very complicated to compute.}
\be
\mathcal{I}_{g=0,s_{2}^{r}=1,m=5,\left(b_{i}=\left(\frac{1}{4},-\frac{1}{4}\right),c_{j}=\left(-\frac{1}{4},\frac{1}{4}\right),e=\frac{3}{4}\right)}^{N=3,k=2}=\frac{\left(1-T^{3}\left(\beta^{-1}\gamma\right)^{3}\right)}{\left(1-T^{2}\right)^{3}\left(1-T^{4}\right)},
\ee
where the fluxes are completely determined by a comparison with \eqref{E:AnsatzPunc}, and assuming \eqref{E:Cmax shift N=2} holds for all $N$. This allows us to generalize \eqref{E:CmaxP N=2} to any $N$. Thus, the change in flux for closing a maximal puncture for any $N$ and $k$ is given by
\be
\label{E:CmaxP}
CmaxP_{flux}=\begin{cases}
\left(\left(b_{i}=\pm\frac{k-1}{\left(N-1\right)k^{2}},b_{j\ne i}=\mp\frac{1}{\left(N-1\right)k^{2}}\right),\vec{0},0\right) & \mbox{each meson vev}\propto\beta_{i}^{\pm1}\\
\left(\vec{0},\left(c_{i}=\pm\frac{k-1}{\left(N-1\right)k^{2}},c_{j\ne i}=\mp\frac{1}{\left(N-1\right)k^{2}}\right),0\right) & \mbox{each meson vev}\propto\gamma_{i}^{\pm1}\\
\left(\vec{0},\vec{0},\mp\frac{k-1}{k^{2}\left(N-1\right)\left(Nk-k-1\right)}\right) & \mbox{each meson vev}\propto t^{\pm1}
\end{cases}
\ee
where again for general $N$, $k(N-1)(Nk-k-1)/2$ mesons get a vev giving a total shift in $e$ of $(k-1)/2k$ as expected.
One can easily see that \eqref{E:CmaxP} reduces to \eqref{E:CmaxP N=2} when setting $N=2$. Also we find in the degenerate case of $k=1$ that there is no shift in flux from closing a maximal puncture to a minimal one as expected. In the even more degenerate case of $k=1$ and $N=2$, one can not distinguish between minimal and maximal punctures, and this formula becomes irrelevant since the closing of punctures is done as if they were all minimal ones. In addition we find the following constraint
\be
\label{E:CmaxP constraints k=2 N=3}
f_{e}\left(3,2\right)+f_{g}\left(3,2\right) & = & 5\nonumber\\
h_{e}\left(2,2\right)+h_{g}\left(2,2\right) & = & 7.
\ee

\subsubsection*{Flipping a maximal puncture and closing it completely}
Flipping a maximal puncture of a free trinion and closing it completely gives a theory with Coulomb index
\be
\mathcal{I}_{g=0,s_{1}^{p}=1,m=1,\left(b_{i},c_{j},e=\frac{3}{4}\right)}^{N=3,k=2} = \frac{1}{\left(1-T^{2}\right)\left(1-T^{4}\right)\left(1-T^{3}(\beta\gamma)^{-3}\right)\left(1-T^{3}(\beta^{-1}\gamma)^{\pm 3}\right)},
\ee
where the fluxes $b_{i}=\left(-\frac{1}{4},\frac{1}{4}\right),c_{j}=\left(-\frac{1}{4},\frac{1}{4}\right)$ were determined by \eqref{E:CmaxP}. Comparing to \eqref{E:AnsatzPunc} we get the following constraints
\be
f_{e}\left(3,2\right)-f_{g}\left(3,2\right) & = & 1\nonumber\\
h_{e}\left(2,2\right)-h_{g}\left(2,2\right) & = & 1.
\ee
These constraints along with \eqref{E:CmaxP constraints k=2 N=3}, \eqref{E:constraints k=N=2} and \eqref{E:constraints k=1} allow us to determine the functions $f_e,\ f_g,\ h_e$ and $h_g$ for any $N,\ k$ and $\ell$ to be
\be
\label{E:he and hg}
f_{e}\left(N,k\right)=N\ ,\ \ \ h_{g}\left(N,k\right)=N-1, \\
h_{e}\left(\ell,k\right)=\ell k\ ,\ \ \ h_{g}\left(\ell,k\right)=\ell k-1.
\ee

\section{The full conjecture}
\label{S:Conj and Apps}

Concentrating all the results we found in section \ref{S:Constraints}, the conjectured formula for the index in the Coulomb limit of a Riemann surface with fluxes and no punctures is
\be
\label{E:conjNoP}
\mathcal{I}_{g,\left(b_{i},c_{j},e\right)}^{N,k} & = & PE\left[\sum_{i,j=1}^{k}\left(-b_{i}+c_{j}+Ne+\left(N-1\right)\left(g-1\right)\right)\beta_{i}^{-N}\gamma_{j}^{N}T^{N}\right]\times\nonumber\\
 & & PE\left[\sum_{\ell=1}^{N-1}\left(\ell ke+\left(\ell k-1\right)\left(g-1\right)\right)T^{\ell k}\right].
\ee
We may as well write the full formula including minimal and maximal punctures
\be
\label{E:conjP}
\mathcal{I}_{g,s_{c}^{r},s_{c}^{\ell},m,\left(b_{i},c_{j},e\right)}^{N,k} & = & PE\left[\sum_{i,j=1}^{k}\left(-b_{i}+c_{j}+N\left(e-\frac{m}{2k}\right)\right)\beta_{i}^{-N}\gamma_{j}^{N}T^{N}\right]\times\nonumber\\
 & & PE\left[\sum_{i,j=1}^{k}\left(\left(N-1\right)\left(g-1+\frac{s_{tot}}{2}\right)+\frac{m}{k}\right)\beta_{i}^{-N}\gamma_{j}^{N}T^{N}\right]\times\nonumber\\
 & & PE\left[\sum_{i,j=1}^{k}\left(\sum_{c,n=1}^{k}\frac{k-2n}{2k}\left(s_{c}^{r}\delta_{i+2-c+n,j}+s_{c}^{\ell}\delta_{i+2-c-n,j}\right)\right)\beta_{i}^{-N}\gamma_{j}^{N}T^{N}\right]\times\nonumber\\
 & & PE\left[\sum_{\ell=1}^{N-1}\left(\ell k\left(e-\frac{m}{2k}\right)+\left(\ell k -1\right)\left(g-1+\frac{s_{tot}}{2}\right)\right)T^{\ell k}\right]\times\nonumber\\
 & & PE\left[\sum_{\ell=1}^{N-1}\left(m-\frac{s_{tot}}{2}\right)T^{\ell k}\right].
\ee
We can have more general maximal punctures corresponding to general permutations of $\beta$ and $\gamma$ symmetries and not just to cyclic permutations. The generalization  of the above to these cases should be straightforward.
These formulas hold for the following flux conventions, with free trinion flux given by
\be
FT_{flux}=\left(0,0,\frac{1}{2k}\right),
\ee
closing minimal puncture flux shift of
\be
\label{E:CminP full}
CminP_{flux}=\begin{cases}
\left(\left(b_{i}=\pm\frac{k-1}{k},b_{j\ne i}=\mp\frac{1}{k}\right),\vec{0},-\frac{1}{2k}\right) & vev\propto\beta_{i}^{\pm 1}\\
\left(\vec{0},\left(c_{i}=\mp\frac{k-1}{k},c_{j\ne i}=\pm\frac{1}{k}\right),-\frac{1}{2k}\right) & vev\propto\gamma_{i}^{\pm 1}
\end{cases},
\ee
and closing maximal puncture flux shift of
\be
\label{E:CmaxP full}
CmaxP_{flux}=\begin{cases}
\left(\left(b_{i}=\pm\frac{k-1}{\left(N-1\right)k^{2}},b_{j\ne i}=\mp\frac{1}{\left(N-1\right)k^{2}}\right),\vec{0},0\right) & \mbox{each meson vev}\propto\beta_{i}^{\pm1}\\
\left(\vec{0},\left(c_{i}=\pm\frac{k-1}{\left(N-1\right)k^{2}},c_{j\ne i}=\mp\frac{1}{\left(N-1\right)k^{2}}\right),0\right) & \mbox{each meson vev}\propto\gamma_{i}^{\pm1}\\
\left(\vec{0},\vec{0},\mp\frac{k-1}{k^{2}\left(N-1\right)\left(Nk-k-1\right)}\right) & \mbox{each meson vev}\propto t^{\pm1}
\end{cases}.
\ee

\section*{Acknowledgments}

We are grateful to Chris Beem, Elli Pomoni, and Gabi Zafrir for  enlightening discussions and very useful comments. The research is  supported by Israel Science Foundation under grant no. 1696/15 and by I-CORE  Program of the Planning and Budgeting Committee.

\vspace{10pt}
\begin{appendix}
\vspace{10pt}
\section{Index notations}
\label{A:pleth}

In this appendix we define a few special functions used throughout this paper and its appendices. The plethystic exponent is defined by
\be
PE \left[f(x,y,...)\right]\triangleq \exp \left[\sum_{n=1}^\infty \frac1{n} f(x^n,y^n,...)\right]
\ee
specifically,
\be
PE [x]=\frac1{1-x} \ , \qquad PE [-x]=1-x.
\ee
The elliptic gamma function is defined as
\be
\Gamma \left( z;p,q\right) \triangleq PE \left[ \frac{z-\frac{pq}{z}}{(1-p)(1-q)}\right] = \prod_{i,j=0}^{\infty} \frac{1-p^{i+1} q^{j+1} z^{-1}}{1-p^{i} q^{j} z}.
\ee
When used, the elliptic gamma function will be given in the implicit form
\be
\Gamma_e(z)=\Gamma \left( z;p,q\right),
\ee
dropping the $p$ and $q$ to shorten the notation. Another shortcut notation is given by
\be
\Gamma_e \left(z^{\pm n} \right)=\Gamma_e \left(z^{n} \right)\Gamma_e \left(z^{-n} \right) \ , \qquad \Gamma_e \left(u z^{\pm n} \right)=\Gamma_e \left(u z^{n} \right)\Gamma_e \left(u z^{-n} \right).
\ee
For 4d ${\cal N}=1$ supersymmetry a contour integral over $T_{N-1}$ of the combined contribution of a vector in the adjoint of $SU(N)$ with the $SU(N)$ Haar measure is given by
\be
\frac{\kappa^{N-1}}{N!} \oint_{T_{N-1}} \prod_{i=1}^{N-1} \frac{dz_i}{2\pi i z_i} \prod_{k\ne \ell} \frac1{\Gamma_e(z_k/z_\ell)}\cdots,
\ee
where the dots denote that it will appear with other functions integrated over as well. $\kappa$ is defined as
\be
\kappa \triangleq (p;p)(q;q),
\ee
where
\be
(a;b) \triangleq \prod_{n=0}^\infty \left( 1-ab^n \right)
\ee
is the q-Pochhammer symbol.

\section{The interacting trinions}
\label{A:intTrin}

In this appendix we write in detail dualities for class ${\cal S}_3$ with $N=2$ and class ${\cal S}_2$ with $N=3$. The dualities are derived from the general procedure of closing punctures detailed in \cite{Gaiotto:2015usa} and correspond to different descriptions of a sphere with two maximal and a collection of minimal punctures. One description has a Lagrangian while the other contains strongly coupled three punctured spheres. We write the duality in terms of the identity of indices. The matter content and the interactions can be read off from such a presentation. The duality also determines the anomalies of the interacting theories which are
compared to the predicted anomalies from 6d given in \cite{Bah:2017gph}. This anomaly comparison serves as an independent check to \eqref{E:CminP full} and \eqref{E:CmaxP full}, where we give the shift in fluxes caused by the closing of maximal and minimal punctures. Since \eqref{E:CminP full} and \eqref{E:CmaxP full} were determined by our conjectured formulas \eqref{E:conjNoP} and \eqref{E:conjP}, this serves as an independent check to these formulas as well. 

The anomalies coming from 6d \cite{Bah:2017gph} of a genus $g$ Riemann surface are given by,

\be
\label{RS 6d anomalies}
 & & TrR'=-\frac{1}{2}(k^{2}-2)(N-1)(2g-2)\,,\qquad Trt=-k^{2}NN_{e}\nonumber\\
 & & TrR'^{3}=\frac{1}{2}(N-1)(k^{2}(N^{2}+N-1)+2)(2g-2)\,,\qquad TrR'^{2}t=\frac{1}{3}k^{2}N(N^{2}-1)N_{e}\nonumber\\
 & & TrR't^{2}=-\frac{1}{6}k^{2}N(N^{2}-1)(2g-2)\,,\qquad Trt^{3}=-k^{2}N^{3}N_{e}\nonumber\\
 & & TrR'^{2}\beta_{i}/\gamma_{i}=-kN(N-1)\left(N_{b_{i}/c_{i}}-N_{b_{k}/c_{k}}\right)\,,\qquad Trt^{2}\beta_{i}/\gamma_{i}=kN^{2}\left(N_{b_{i}/c_{i}}-N_{b_{k}/c_{k}}\right)\nonumber\\
 & & TrR'\beta_{i}^{2}/\gamma_{i}^{2}=-kN^{2}(N-1)(2g-2)\,,\qquad TrR'\left(\beta_{i}\beta_{j}/\gamma_{i}\gamma_{j}\right)=-\frac{1}{2}kN^{2}(N-1)(2g-2)\nonumber\\
 & & Trt\beta_{i}^{2}=-kN^{2}\left(2NN_{e}-\left(N_{b_{i}}+N_{b_{k}}\right)\right)\,,\qquad Trt\gamma_{i}^{2}=-kN^{2}\left(2NN_{e}+\left(N_{c_{i}}+N_{c_{k}}\right)\right)\nonumber\\
 & & Trt\beta_{i}\beta_{j}=-kN^{2}\left(NN_{e}-N_{b_{k}}\right)\,,\qquad Trt\gamma_{i}\gamma_{j}=-kN^{2}\left(NN_{e}+N_{c_{k}}\right)\nonumber\\
 & & Tr\beta_{i}/\gamma_{i}=kN\left(N_{b_{i}/c_{i}}-N_{b_{k}/c_{k}}\right)\,,\qquad Tr\beta_{i}^{3}/\gamma_{i}^{3}=N^{2}\left(kN+6(N-1)\right)\left(N_{b_{i}/c_{i}}-N_{b_{k}/c_{k}}\right)\nonumber\\
 & & Tr\beta_{i}^{2}\beta_{j}=2N^{2}(N-1)\left(N_{b_{i}}+N_{b_{j}}-2N_{b_{k}}\right)+kN^{3}\left(N_{e}-N_{b_{k}}\right)\nonumber\\
 & & Tr\gamma_{i}^{2}\gamma_{j}=2N^{2}(N-1)\left(N_{c_{i}}+N_{c_{j}}-2N_{c_{k}}\right)-kN^{3}\left(N_{e}+N_{c_{k}}\right)\nonumber\\
 & & Tr\beta_{i}\beta_{j}\beta_{\ell}=N^{2}(N-1)\left(N_{b_{i}}+N_{b_{j}}+N_{b_{\ell}}-3N_{b_{k}}\right)+kN^{3}\left(N_{e}-N_{b_{k}}\right)\nonumber\\
 & & Tr\gamma_{i}\gamma_{j}\gamma_{\ell}=N^{2}(N-1)\left(N_{c_{i}}+N_{c_{j}}+N_{c_{\ell}}-3N_{c_{k}}\right)-kN^{3}\left(N_{e}+N_{c_{k}}\right)\nonumber\\
 & & Tr\left(\beta_{i}^{2}\gamma_{j}/\gamma_{i}^{2}\beta_{j}\right)=2N^{2}\left(N_{c_{j}/b_{j}}-N_{c_{k}/b_{k}}\right)\nonumber\\
 & & Tr\left(\beta_{i}\beta_{j}\gamma_{\ell}/\gamma_{i}\gamma_{j}\beta_{\ell}\right)=N^{2}\left(N_{c_{\ell}/b_{\ell}}-N_{c_{k}/b_{k}}\right)
\ee
where $N_{b_i}$, $N_{c_j}$ and $N_e$ are the $b_i$, $c_j$ and $e$ components of flux, respectively, the slashes appearing in some of the formulas are correlated, and the anomalies not written vanish for all parameters.

\subsection*{N=2 k=3 interacting trinion}

\begin{figure}[t]
        \centering
        \includegraphics[scale=0.3]{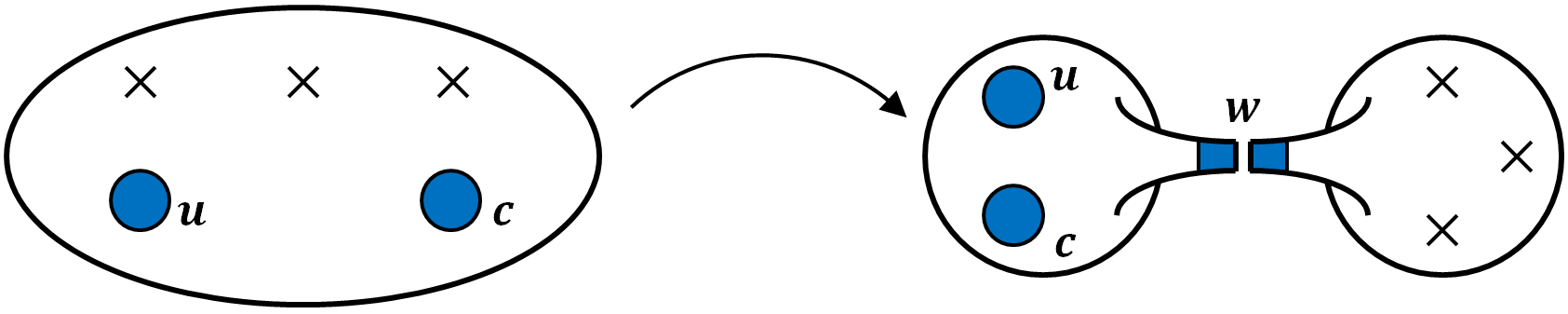}
        \caption{The duality frame used for $N=2,\ k=3$. On the left side of the duality, three free trinions glued together with two maximal punctures of color $c=1$ (blue) and three minimal punctures. On the right side of the duality, a tail with one maximal puncture of color $c=1$ and three minimal punctures glued to an interacting trinion with three maximal punctures of color $c=1$.}
        \label{F:ITduality N=2 k=3}
\end{figure}

The full superconformal index of the tail we used in section \ref{S:Constraints} can be used to find the interacting trinion, as shown in figure \ref{F:ITduality N=2 k=3}, with three maximal punctures of color $c=1$, one of orientation $r$ ($\mathbf{c}$) and two of orientation $\ell$ ($\mathbf{u}$ and $\mathbf{w}$). 
We denote the index of this interacting trinion as $\mathcal{I}_{IT\,\mathbf{u}\mathbf{w}}^{\mathbf{c}}$ and it is given as a duality between the tail glued to the interacting trinion
\be
\mathcal{I}_{\mathbf{u}\alpha\rho\delta}^{\mathbf{c}} & = & \Gamma_{e}\left(t\gamma_{2}^{-2}\alpha^{2}\right)\Gamma_{e}\left(t\gamma_{1}^{2}\gamma_{2}^{2}\alpha^{2}\right)\Gamma_{e}\left(t\beta_{2}^{2}\alpha^{-2}\right)\Gamma_{e}\left(t\beta_{1}^{-2}\beta_{2}^{-2}\alpha^{-2}\right)\Gamma_{e}\left(t\beta_{1}^{-2}\beta_{2}^{-2}\rho^{-2}\right)\times\nonumber\\
 & & \Gamma_{e}\left(t\gamma_{2}^{-2}\delta^{2}\right)\Gamma_{e}\left(t\gamma_{1}^{2}\gamma_{2}^{2}\delta^{2}\right)\Gamma_{e}\left(t\beta_{2}^{2}\delta^{-2}\right)\Gamma_{e}\left(t\beta_{1}^{-2}\beta_{2}^{-2}\delta^{-2}\right)\Gamma_{e}\left(t\gamma_{2}^{-2}\rho^{2}\right)\Gamma_{e}\left(pqt\gamma_{1}^{2}\beta_{1}^{-2}\right)\times\nonumber\\
 & & \kappa\oint\frac{dz_{1}}{4\pi iz_{1}}\kappa^{2}\oint\frac{dw_{1}}{4\pi iw_{1}}\oint\frac{dw_{2}}{4\pi iw_{2}}\mathcal{I}_{IT\,\mathbf{u}\left\{ w_{1},w_{2},\rho\alpha\delta\right\} }^{\mathbf{c}}\Gamma_{e}\left(\frac{pq}{t^{1/2}}\frac{\left(\beta_{2}^{-1}\gamma_{1}\gamma_{2}\alpha\delta\right)^{\pm1}}{\beta_{1}\beta_{2}\gamma_{2}}z_{1}^{\pm1}\right)\times\nonumber\\
 & & \frac{\Gamma_{e}\left(\frac{pq}{t}\gamma_{2}\beta_{1}^{-1}w_{1}^{\pm1}\left(\rho\alpha\delta\right)^{-1}\right)\Gamma_{e}\left(\frac{pq}{t}\beta_{1}\beta_{2}\gamma_{1}w_{2}^{\pm1}\rho\alpha\delta\right)\Gamma_{e}\left(\frac{pq}{t}\beta_{2}^{-1}\gamma_{1}^{-1}\gamma_{2}^{-1}w_{1}^{\pm1}w_{2}^{\pm1}\right)}{\Gamma_{e}\left(z_{1}^{\pm2}\right)\Gamma_{e}\left(w_{1}^{\pm2}\right)\Gamma_{e}\left(w_{2}^{\pm2}\right)}\times\nonumber\\
 & & \Gamma_{e}\left(t^{-\frac{1}{2}}z_{1}^{\pm1}\gamma_{1}^{-1}\beta_{1}\left(\alpha\delta^{-1}\right)^{\pm1}\right)\Gamma_{e}\left(t^{\frac{1}{2}}z_{1}^{\pm1}w_{1}^{\pm1}\gamma_{1}\gamma_{2}\rho\right)\Gamma_{e}\left(t^{\frac{1}{2}}z_{1}^{\pm1}w_{2}^{\pm1}\beta_{2}\rho^{-1}\right)\times\nonumber\\
 & & \Gamma_{e}\left(w_{2}^{\pm1}\gamma_{1}^{-1}\beta_{1}\beta_{2}\rho\left(\alpha\delta\right)^{-1}\right)\Gamma_{e}\left(w_{1}^{\pm1}\gamma_{2}\beta_{1}\rho^{-1}\alpha\delta\right),
\ee
and the theory of three glued free trinions
\be
\label{E:Orbifold N=2 k=3}
\mathcal{I}_{\mathbf{u}\alpha\rho\delta}^{\mathbf{c}} & = & \kappa^{3}\oint\frac{dz_{1}}{4\pi iz_{1}}\oint\frac{dz_{2}}{4\pi iz_{2}}\oint\frac{dz_{3}}{4\pi iz_{3}}\kappa^{3}\oint\frac{dw_{1}}{4\pi iw_{1}}\oint\frac{dw_{2}}{4\pi iw_{2}}\oint\frac{dw_{3}}{4\pi iw_{3}}\times\nonumber\\ 
& & \frac{\Gamma_{e}\left(\frac{pq}{t}\beta_{2}^{-1}\gamma_{2}z_{1}^{\pm1}z_{2}^{\pm1}\right)\Gamma_{e}\left(\frac{pq}{t}\beta_{1}\beta_{2}\gamma_{1}^{-1}\gamma_{2}^{-1}z_{2}^{\pm1}z_{3}^{\pm1}\right)\Gamma_{e}\left(\frac{pq}{t}\beta_{1}^{-1}\gamma_{1}z_{3}^{\pm1}z_{1}^{\pm1}\right)}{\Gamma_{e}\left(z_{1}^{\pm2}\right)\Gamma_{e}\left(z_{2}^{\pm2}\right)\Gamma_{e}\left(z_{3}^{\pm2}\right)}\times\nonumber\\ 
& & \frac{\Gamma_{e}\left(\frac{pq}{t}\gamma_{1}\beta_{2}^{-1}w_{1}^{\pm1}w_{2}^{\pm1}\right)\Gamma_{e}\left(\frac{pq}{t}\beta_{1}\beta_{2}\gamma_{2}w_{2}^{\pm1}w_{3}^{\pm1}\right)\Gamma_{e}\left(\frac{pq}{t}\beta_{1}^{-1}\gamma_{1}^{-1}\gamma_{2}^{-1}w_{3}^{\pm1}w_{1}^{\pm1}\right)}{\Gamma_{e}\left(w_{1}^{\pm2}\right)\Gamma_{e}\left(w_{2}^{\pm2}\right)\Gamma_{e}\left(w_{3}^{\pm2}\right)}\times\nonumber\\ 
& & \Gamma_{e}\left(t^{\frac{1}{2}}z_{3}^{\pm1}u_{1}^{\pm1}\gamma_{1}^{-1}\alpha\right)\Gamma_{e}\left(t^{\frac{1}{2}}z_{1}^{\pm1}u_{1}^{\pm1}\beta_{1}\alpha^{-1}\right)\Gamma_{e}\left(t^{\frac{1}{2}}z_{1}^{\pm1}u_{2}^{\pm1}\gamma_{2}^{-1}\alpha\right)\times\nonumber\\ 
& & \Gamma_{e}\left(t^{\frac{1}{2}}z_{2}^{\pm1}u_{2}^{\pm1}\beta_{2}\alpha^{-1}\right)\Gamma_{e}\left(t^{\frac{1}{2}}z_{2}^{\pm1}u_{3}^{\pm1}\gamma_{1}\gamma_{2}\alpha\right)\Gamma_{e}\left(t^{\frac{1}{2}}z_{3}^{\pm1}u_{3}^{\pm1}\beta_{1}^{-1}\beta_{2}^{-1}\alpha^{-1}\right)\times\nonumber\\ 
& & \Gamma_{e}\left(t^{\frac{1}{2}}z_{1}^{\pm1}w_{1}^{\pm1}\gamma_{1}^{-1}\rho\right)\Gamma_{e}\left(t^{\frac{1}{2}}z_{1}^{\pm1}w_{2}^{\pm1}\beta_{2}\rho^{-1}\right)\Gamma_{e}\left(t^{\frac{1}{2}}z_{2}^{\pm1}w_{2}^{\pm1}\gamma_{2}^{-1}\rho\right)\times\nonumber\\ 
& & \Gamma_{e}\left(t^{\frac{1}{2}}z_{2}^{\pm1}w_{3}^{\pm1}\beta_{1}^{-1}\beta_{2}^{-1}\rho^{-1}\right)\Gamma_{e}\left(t^{\frac{1}{2}}z_{3}^{\pm1}w_{3}^{\pm1}\gamma_{1}\gamma_{2}\rho\right)\Gamma_{e}\left(t^{\frac{1}{2}}z_{3}^{\pm1}w_{1}^{\pm1}\beta_{1}\rho^{-1}\right)\times\nonumber\\ 
& & \Gamma_{e}\left(t^{\frac{1}{2}}w_{2}^{\pm1}c_{2}^{\pm1}\gamma_{1}^{-1}\delta\right)\Gamma_{e}\left(t^{\frac{1}{2}}w_{2}^{\pm1}c_{3}^{\pm1}\beta_{1}^{-1}\beta_{2}^{-1}\delta^{-1}\right)\Gamma_{e}\left(t^{\frac{1}{2}}w_{3}^{\pm1}c_{3}^{\pm1}\gamma_{2}^{-1}\delta\right)\times\nonumber\\ 
& & \Gamma_{e}\left(t^{\frac{1}{2}}w_{3}^{\pm1}c_{1}^{\pm1}\beta_{1}\delta^{-1}\right)\Gamma_{e}\left(t^{\frac{1}{2}}w_{1}^{\pm1}c_{1}^{\pm1}\gamma_{1}\gamma_{2}\delta\right)\Gamma_{e}\left(t^{\frac{1}{2}}w_{1}^{\pm1}c_{2}^{\pm1}\beta_{2}\delta^{-1}\right),
\ee
and its fluxes are ${\cal F} = \left(\left(\frac{1}{3},0,-\frac{1}{3}\right),\left(-\frac{1}{3},\frac{1}{3},0\right),\frac{1}{2}\right)$.

In order to compare to the anomalies coming from $6d$ we need a closed Riemann surface, thus we need to glue the aforementioned interacting trinion to another interacting trinion with opposite puncture orientations. We denote this trinion as $\mathcal{I}_{IT\,\mathbf{u}}^{\mathbf{c}\mathbf{z}}$ (superscript and subscript fugacities are for right and left orientation, respectively) and give its index as a duality between the tail glued to the interacting trinion
\be
\mathcal{I}_{\mathbf{u}\alpha\rho\delta}^{\mathbf{c}} & = & \Gamma_{e}\left(t\gamma_{2}^{-2}\alpha^{2}\right)\Gamma_{e}\left(t\gamma_{1}^{2}\gamma_{2}^{2}\alpha^{2}\right)\Gamma_{e}\left(t\beta_{1}^{2}\alpha^{-2}\right)\Gamma_{e}\left(t\beta_{1}^{-2}\beta_{2}^{-2}\alpha^{-2}\right)\Gamma_{e}\left(t\beta_{1}^{-2}\beta_{2}^{-2}\rho^{-2}\right)\times\nonumber\\ 
& & \Gamma_{e}\left(t\gamma_{2}^{-2}\delta^{2}\right)\Gamma_{e}\left(t\gamma_{1}^{2}\gamma_{2}^{2}\delta^{2}\right)\Gamma_{e}\left(t\beta_{1}^{2}\delta^{-2}\right)\Gamma_{e}\left(t\beta_{1}^{-2}\beta_{2}^{-2}\delta^{-2}\right)\Gamma_{e}\left(t\gamma_{1}^{2}\gamma_{2}^{2}\rho^{2}\right)\Gamma_{e}\left(pqt\gamma_{1}^{2}\beta_{2}^{-2}\right)\times\nonumber\\ 
& & \kappa^{2}\oint\frac{dz_{1}}{4\pi iz_{1}}\oint\frac{dz_{2}}{4\pi iz_{2}}\kappa\oint\frac{dw_{1}}{4\pi iw_{1}}\mathcal{I}_{IT\,\mathbf{u}}^{\mathbf{c}\left\{ z_{1},z_{2},\rho\epsilon\delta\right\} }\Gamma_{e}\left(\frac{pq}{t^{1/2}}\frac{\gamma_{1}\gamma_{2}}{\beta_{1}\beta_{2}}\left(\beta_{1}^{-1}\gamma_{2}^{-1}\alpha\delta\right)^{\pm1}w_{1}^{\pm1}\right)\times\nonumber\\ 
& & \frac{\Gamma_{e}\left(\frac{pq}{t}\gamma_{1}\beta_{1}\beta_{2}z_{1}^{\pm1}\rho\alpha\delta\right)\Gamma_{e}\left(\frac{pq}{t}\beta_{2}^{-1}\gamma_{1}^{-1}\gamma_{2}^{-1}z_{2}^{\pm1}\left(\rho\alpha\delta\right)^{-1}\right)\Gamma_{e}\left(\frac{pq}{t}\beta_{1}^{-1}\gamma_{2}z_{1}^{\pm1}z_{2}^{\pm1}\right)}{\Gamma_{e}\left(w_{1}^{\pm2}\right)\Gamma_{e}\left(z_{1}^{\pm2}\right)\Gamma_{e}\left(z_{2}^{\pm2}\right)}\times\nonumber\\ 
& & \Gamma_{e}\left(t^{-\frac{1}{2}}w_{1}^{\pm1}\gamma_{1}^{-1}\beta_{2}\left(\alpha\delta^{-1}\right)^{\pm1}\right)\Gamma_{e}\left(t^{\frac{1}{2}}z_{2}^{\pm1}w_{1}^{\pm1}\gamma_{2}^{-1}\rho\right)\Gamma_{e}\left(t^{\frac{1}{2}}z_{1}^{\pm1}w_{1}^{\pm1}\beta_{1}\rho^{-1}\right)\times\nonumber\\ 
& & \Gamma_{e}\left(z_{1}^{\pm1}\gamma_{1}^{-1}\beta_{1}\beta_{2}\rho\left(\alpha\delta\right)^{-1}\right)\Gamma_{e}\left(z_{2}^{\pm1}\gamma_{1}^{-1}\gamma_{2}^{-1}\beta_{2}\rho^{-1}\alpha\delta\right),
\ee
and the theory of three free trinions glued together given in \eqref{E:Orbifold N=2 k=3}. This interacting trinion has a flux of ${\cal F} = \left(\left(0,\frac{1}{3},-\frac{1}{3}\right),\left(-\frac{1}{3},0,\frac{1}{3}\right),\frac{1}{2}\right)$.

Now we can compute the anomalies for the two interacting trinions along with the anomalies coming from three color $c=1$ $\Phi$-gluings of the form
\be
\left(\prod_{n=1}^{3}\kappa\oint\frac{dz_{n}}{4\pi iz_{n}}\right)\Gamma_{e}\left(\frac{pq}{t}\frac{z_{1}^{\pm1}z_{2}^{\pm1}}{\beta_{2}\gamma_{1}\gamma_{2}}\right)\frac{\Gamma_{e}\left(\frac{pq}{t}\beta_{1}\beta_{2}\gamma_{1}z_{2}^{\pm1}z_{3}^{\pm1}\right)\Gamma_{e}\left(\frac{pq}{t}\beta_{1}^{-1}\gamma_{2}z_{3}^{\pm1}z_{1}^{\pm1}\right)}{\Gamma_{e}\left(z_{1}^{\pm2}\right)\Gamma_{e}\left(z_{2}^{\pm2}\right)\Gamma_{e}\left(z_{3}^{\pm2}\right)},
\ee
and compare to the predicted anomalies from 6d given in \eqref{RS 6d anomalies} with genus $g=2$, and remember that we need to compute it using the six dimensional symmetry which implies that in field theory computation we take $t$ to $ t (p q)^{\frac12}$. The results are shown in table \ref{T:IT anomalies N=2 k=3}.
\begin{table}[hbt]
	\begin{center}
		\begin{tabular}{ | c || p{9.5cm} | p{2.5cm} | c | }
  		\hline			
  		 & \ \ \ \ \ \ \ \ \ \ \ \ \ \ \ \ \ \ \ \ \ \ \ \ \ \ 4d calculation & \ 6d prediction & $A$\\ \hline\hline
  		$TrR'$ & $[3\times6+(\frac{1}{2}-1)72]-[(3-1)+3\times3+8(\frac{1}{2}-1)+4(\frac{3}{2}-1)+4(-\frac{1}{2}-1)+4(-1)]+[3\times6+(\frac{1}{2}-1)72]-[(3-1)+3\times3+4(-\frac{1}{2}-1)+4(-1)+4(\frac{3}{2}-1)+8(\frac{1}{2}-1)]+3\times(3\times3)$ & $-\frac12(3^2-2)(2-1)\cdot(2\cdot2-2)$ & $-7$ \\ \hline
  		$Trt$ & $[(-1)24+(\frac{1}{2})72]-[11(1)+8(-1)+8(\frac{1}{2})+8(-\frac{1}{2})]+[(-1)24+(\frac{1}{2})72]-[11(1)+8(-1)+8(\frac{1}{2})+8(-\frac{1}{2})]+3\times12(-1)$ & $-\frac12 (2 \cdot 3^2\cdot 2 \cdot 1)$ & $-18$ \\ \hline
  		$Tr\beta_1$ & $-[2(2)+5(-2)+4(-1)+8(1)]-[4(-2)+10(1)+6(-1)]$ & $3\cdot2\left(\frac1{3}-(-\frac{2}{3})\right)$ & $6$ \\ \hline
  		$Tr\beta_2$ & $-[2(2)+5(-2)+4(-1)+8(1)]-[4(-2)+6(-1)+10(1)]$ & $3\cdot2\left(\frac1{3}-(-\frac{2}{3})\right)$ & $6$ \\ \hline
  		$Tr\gamma_1$ & $-[3(2)+12(-1)+8(1)]-[4(2)+6(1)+10(-1)]$ & $3\cdot2\left(-\frac{2}{3}-\frac1{3}\right)$ & $-6$ \\ \hline
  		$Tr\gamma_2$ & $-[2(2)+5(-2)+8(1)+4(-1)]-[2(-2)+5(2)+4(1)+8(-1)]$ & $3\cdot2\left(-\frac1{3}+\frac1{3}\right)$ & $0$ \\
  		\hline 
		\end{tabular}
	\end{center}
	\caption{Comparison of linear anomalies of the genus two Riemann surface built out of two interacting trinions, and the 6d predictions. The rest of the anomalies match as well, but are not shown for abbreviation. The predictions from 6d are calculated using the combined fluxes of both interacting trinions, with a total flux of ${\cal F}=\left( b_i=(\frac1{3},\frac1{3},-\frac{2}{3}),c_j=(-\frac{2}{3},\frac1{3},\frac1{3}),e=1\right)$.}
	\label{T:IT anomalies N=2 k=3}
\end{table}
This give us another validation of our results.

\subsection*{N=3 k=2 interacting trinion}

\begin{figure}[t]
        \centering
        \includegraphics[scale=0.3]{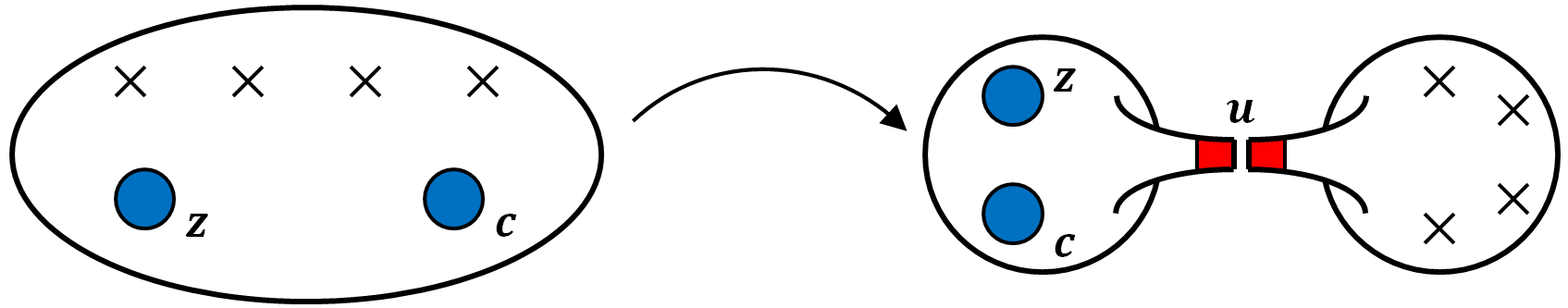}
        \caption{The duality frame used for $N=3,\ k=2$. On the left side of the duality, four free trinions glued together with two maximal punctures of color $c=1$ (blue) and four minimal punctures. On the right side of the duality, a tail with one maximal puncture of color $c=2$ (red) and four minimal punctures glued to an interacting trinion with three maximal punctures two of color $c=1$ and one of color $c=2$.}
        \label{F:ITduality N=3 k=2}
\end{figure}

In the same manner as in the case of $N=2,\ k=3$ we find the interacting trinion, as shown in figure \ref{F:ITduality N=3 k=2}, with two maximal punctures of color $c=1$ of both orientations ($\mathbf{c}$ of orientation $r$ and $\mathbf{z}$ of orientation $\ell$), and one maximal puncture of color $c=2$ with orientation $\ell$ ($\mathbf{u}$). 
We similarly denote the index of this interacting trinion as $\mathcal{I}_{IT\,\mathbf{z}\mathbf{u}}^{\mathbf{c}}$ and it is given as a duality between the tail glued to the interacting trinion
\be
\mathcal{I}_{\mathbf{z}\alpha\delta\epsilon\rho}^{\mathbf{c}} & = & \Gamma_{e}\left(t^{3/2}\beta^{3}\delta^{-3}\right)\Gamma_{e}\left(t^{3/2}\gamma^{-3}\delta^{3}\right)\Gamma_{e}\left(t^{3/2}\beta^{3}\alpha^{-3}\right)\Gamma_{e}\left(t^{3/2}\gamma^{-3}\alpha^{3}\right)\times\nonumber\\
 & & \Gamma_{e}\left(pqt\beta^{3}\gamma^{-3}\right)\kappa\oint\frac{du_{1}}{4\pi iu_{1}}\frac{\kappa^{2}}{6}\oint\frac{du_{2}^{(1)}}{2\pi iu_{2}^{(1)}}\oint\frac{du_{2}^{(2)}}{2\pi iu_{2}^{(2)}}\kappa^{3}\oint\frac{dw}{4\pi iw}\oint\frac{dv_{1}}{4\pi iv_{1}}\oint\frac{dv_{2}}{4\pi iv_{2}}\times\nonumber\\
 & & \mathcal{I}_{IT\mathbf{z}\left\{ \rho\epsilon\alpha\delta,u_{1}\left(\rho\epsilon\alpha\delta\right)^{-1/2},u_{2}^{(1)},u_{2}^{(2)}\right\} }^{\mathbf{c}}\frac{\Gamma_{e}\left(\frac{pq}{t}\left(t^{1/2}\beta^{3/2}\gamma^{-3/2}\right)^{\pm1}v_{1}^{\pm1}v_{2}^{\pm1}\right)}{\Gamma_{e}\left(v_{1}^{\pm2}\right)\Gamma_{e}\left(v_{2}^{\pm2}\right)}\times\nonumber\\
 & & \frac{\Gamma_{e}\left(\frac{pq}{t}w^{\pm1}\left(\beta\gamma\alpha^{-1}\delta^{-1}\right)^{\pm3/2}\right)\prod_{b=1}^{3}\Gamma_{e}\left(\frac{pq}{t}u_{1}^{\pm1}\left(\left(\beta\gamma\right)\left(\left(\rho\epsilon\alpha\delta\right)^{1/2}u_{2}^{(b)}\right)^{-1}\right)^{\pm1}\right)}{\Gamma_{e}\left(w^{\pm2}\right)\Gamma_{e}\left(u_{1}^{\pm2}\right)\prod_{i=1}^{3}\prod_{j=i+1}^{3}\Gamma_{e}\left(\left(u_{2}^{(i)}\left(u_{2}^{(j)}\right)^{-1}\right)^{\pm1}\right)}\times\nonumber\\
 & & \Gamma_{e}\left(t^{-1/2}\beta^{-3/2}\gamma^{3/2}w^{\pm1}\left(\alpha^{-1}\delta\right)^{\pm3/2}\right)\Gamma_{e}\left(t^{3/4}\gamma^{-3/2}\epsilon^{3/2}w^{\pm1}v_{1}^{\pm1}\right)\times\nonumber\\
 & & \Gamma_{e}\left(t^{3/4}\beta^{3/2}\epsilon^{-3/2}w^{\pm1}v_{2}^{\pm1}\right)\Gamma_{e}\left(t^{3/4}\beta^{3/2}\rho^{-3/2}v_{1}^{\pm1}u_{1}^{\pm1}\right)\Gamma_{e}\left(t^{3/4}\gamma^{-3/2}\rho^{3/2}v_{2}^{\pm1}u_{1}^{\pm1}\right)\times\nonumber\\
 & & \Gamma_{e}\left(t^{1/4}\left(\gamma\epsilon\left(\alpha\delta\right)^{-1}\right)^{3/2}v_{2}^{\pm1}\right)\Gamma_{e}\left(t^{1/4}\left(\left(\beta\epsilon\right)^{-1}\alpha\delta\right)^{3/2}v_{1}^{\pm1}\right)\times\nonumber\\
 & & \prod_{b=1}^{3}\Gamma_{e}\left(t^{1/4}\gamma\rho\left(\beta\epsilon\alpha\delta\right)^{-1/2}v_{1}^{\pm1}\left(u_{2}^{(b)}\right)^{-1}\right)\Gamma_{e}\left(t^{1/4}\beta^{-1}\rho^{-1}\left(\gamma\epsilon\alpha\delta\right)^{1/2}v_{2}^{\pm1}u_{2}^{(b)}\right),
\ee
and the theory of four glued free trinions
\be
\label{E:Orbifold N=3 k=2}
\mathcal{I}_{\mathbf{z}\alpha\delta\epsilon\rho}^{\mathbf{c}} & = & \left(\frac{\kappa^{2}}{6}\right)^{6}\prod_{i,j=1}^{2}\oint\frac{dw_{i}^{(j)}}{2\pi iw_{i}^{(j)}}\oint\frac{dv_{i}^{(j)}}{2\pi iv_{i}^{(j)}}\oint\frac{du_{i}^{(j)}}{2\pi iu_{i}^{(j)}}\prod_{a,b=1}^{3}\Gamma_{e}\left(\frac{pq}{t}\left(\beta\gamma w_{1}^{(a)}\left(w_{2}^{(b)}\right)^{-1}\right)^{\pm1}\right)\times\nonumber\\
 & & \frac{\Gamma_{e}\left(\frac{pq}{t}\left(\beta^{-1}\gamma v_{1}^{(a)}\left(v_{2}^{(b)}\right)^{-1}\right)^{\pm1}\right)\prod_{a,b=1}^{3}\Gamma_{e}\left(\frac{pq}{t}\left(\beta\gamma u_{1}^{(a)}\left(u_{2}^{(b)}\right)^{-1}\right)^{\pm1}\right)}{\prod_{\ell=1}^{2}\prod_{i,j=1,i\ne j}^{3}\Gamma_{e}\left(w_{\ell}^{(i)}\left(w_{\ell}^{(j)}\right)^{-1}\right)\Gamma_{e}\left(v_{\ell}^{(i)}\left(v_{\ell}^{(j)}\right)^{-1}\right)\Gamma_{e}\left(u_{\ell}^{(i)}\left(u_{\ell}^{(j)}\right)^{-1}\right)}\times\nonumber\\
 & & \mathcal{I}_{FT,\mathbf{z}\alpha}^{\mathbf{w}}\left(\gamma,\beta\right)\mathcal{I}_{FT,\mathbf{w}\delta}^{\mathbf{v}}\left(\gamma,\beta^{-1}\right)\mathcal{I}_{FT,\mathbf{v}\epsilon}^{\mathbf{u}}\left(\gamma,\beta\right)\mathcal{I}_{FT,\mathbf{u}\rho}^{\mathbf{c}}\left(\gamma,\beta^{-1}\right),
\ee
with
\be
\mathcal{I}_{FT,\mathbf{z}\alpha}^{\mathbf{w}}\left(\gamma,\beta\right)&=&\prod_{a,b=1}^{3}\Gamma_{e}\left(t^{\frac{1}{2}}\gamma\alpha z_{1}^{(a)}\left(w_{2}^{(b)}\right)^{-1}\right)\Gamma_{e}\left(t^{\frac{1}{2}}\beta\alpha^{-1}\left(z_{1}^{(a)}\right)^{-1}w_{1}^{(b)}\right)\times\nonumber\\
 & & \prod_{a,b=1}^{3}\Gamma_{e}\left(t^{\frac{1}{2}}\gamma^{-1}\alpha z_{2}^{(a)}\left(w_{1}^{(b)}\right)^{-1}\right)\Gamma_{e}\left(t^{\frac{1}{2}}\beta^{-1}\alpha^{-1}\left(z_{2}^{(a)}\right)^{-1}w_{2}^{(b)}\right),
\ee
and its fluxes are ${\cal F}=\left(\left(-\frac{1}{4},\frac{1}{4}\right),\left(\frac{1}{4},-\frac{1}{4}\right),\frac{1}{2}\right)$.

In the special case of $k=2$, there is an interacting trinion with  opposite orientation maximal punctures, same fluxes, and with the same index up to exchanges of fundamental and anti-fundamental representation of some of the chiral multiplets under the gauge group. This obviously leads to equal contribution from both trinions to all of the anomalies. We denote this trinion as ${\cal I}_{IT\ z}^{cw}$, and we write its index for completion as a duality between the tail glued to the interacting trinion
\be
\mathcal{I}_{\mathbf{z}\alpha\delta\epsilon\rho}^{\mathbf{c}} & = & \Gamma_{e}\left(t^{3/2}\beta^{3}\delta^{-3}\right)\Gamma_{e}\left(t^{3/2}\gamma^{-3}\delta^{3}\right)\Gamma_{e}\left(t^{3/2}\beta^{3}\alpha^{-3}\right)\Gamma_{e}\left(t^{3/2}\gamma^{-3}\alpha^{3}\right)\times\nonumber\\
 & & \Gamma_{e}\left(pqt\beta^{3}\gamma^{-3}\right)\kappa\oint\frac{dw_{2}}{4\pi iw_{2}}\frac{\kappa^{2}}{6}\oint\frac{dw_{1}^{(1)}}{2\pi iw_{1}^{(1)}}\oint\frac{dw_{1}^{(2)}}{2\pi iw_{1}^{(2)}}\kappa^{3}\oint\frac{du}{4\pi iu}\oint\frac{dv_{1}}{4\pi iv_{1}}\oint\frac{dv_{2}}{4\pi iv_{2}}\times\nonumber\\
 & & \mathcal{I}_{IT\mathbf{z}}^{\mathbf{c}\left\{ w_{1}^{(1)},w_{1}^{(2)},\left(\rho\epsilon\alpha\delta\right)^{-1},w_{2}\left(\rho\epsilon\alpha\delta\right)^{1/2}\right\} }\frac{\Gamma_{e}\left(\frac{pq}{t}\left(t^{1/2}\beta^{3/2}\gamma^{-3/2}\right)^{\pm1}v_{1}^{\pm1}v_{2}^{\pm1}\right)}{\Gamma_{e}\left(v_{1}^{\pm2}\right)\Gamma_{e}\left(v_{2}^{\pm2}\right)}\times\nonumber\\
 & & \frac{\Gamma_{e}\left(\frac{pq}{t}u^{\pm1}\left(\beta\gamma\alpha^{-1}\delta^{-1}\right)^{\pm3/2}\right)\prod_{b=1}^{3}\Gamma_{e}\left(\frac{pq}{t}w_{2}^{\pm1}\left(\beta\gamma\left(\rho\epsilon\alpha\delta\right)^{-1/2}w_{1}^{(b)}\right)^{\pm1}\right)}{\Gamma_{e}\left(w^{\pm2}\right)\Gamma_{e}\left(u_{1}^{\pm2}\right)\prod_{i,j=1,i\ne j}^{3}\Gamma_{e}\left(u_{2}^{(i)}\left(u_{2}^{(j)}\right)^{-1}\right)}\times\nonumber\\
 & & \Gamma_{e}\left(t^{-1/2}\beta^{-3/2}\gamma^{3/2}u^{\pm1}\left(\alpha^{-1}\delta\right)^{\pm3/2}\right)\Gamma_{e}\left(t^{3/4}\gamma^{-3/2}\epsilon^{3/2}u^{\pm1}v_{2}^{\pm1}\right)\times\nonumber\\
 & & \Gamma_{e}\left(t^{3/4}\beta^{3/2}\epsilon^{-3/2}u^{\pm1}v_{1}^{\pm1}\right)\Gamma_{e}\left(t^{3/4}\beta^{3/2}\rho^{-3/2}v_{2}^{\pm1}w_{2}^{\pm1}\right)\Gamma_{e}\left(t^{3/4}\gamma^{-3/2}\rho^{3/2}v_{1}^{\pm1}w_{2}^{\pm1}\right)\times\nonumber\\
 & & \Gamma_{e}\left(t^{1/4}\left(\gamma\epsilon\left(\alpha\delta\right)^{-1}\right)^{3/2}v_{1}^{\pm1}\right)\Gamma_{e}\left(t^{1/4}\left(\left(\beta\epsilon\right)^{-1}\alpha\delta\right)^{3/2}v_{2}^{\pm1}\right)\times\nonumber\\
 & & \prod_{b=1}^{3}\Gamma_{e}\left(t^{1/4}\gamma\rho\left(\beta\epsilon\alpha\delta\right)^{-1/2}v_{2}^{\pm1}w_{1}^{(b)}\right)\Gamma_{e}\left(t^{1/4}\beta^{-1}\rho^{-1}\left(\gamma\epsilon\alpha\delta\right)^{1/2}v_{1}^{\pm1}\left(w_{1}^{(b)}\right)^{-1}\right),
\ee
and the theory of four glued free trinions \eqref{E:Orbifold N=3 k=2}.

Using these dualities, and the gluing
\be
\left(\frac{\kappa^{2}}{6}\right)^{2}\left(\prod_{i,j=1}^{2}\oint\frac{dz_{i}^{(j)}}{2\pi iz_{i}^{(j)}}\right)\frac{\prod_{a,b=1}^{3}\Gamma_{e}\left(\frac{pq}{t}\left(\beta\gamma z_{1}^{(a)}(z_{2}^{(b)})^{-1}\right)^{\pm1}\right)}{\prod_{i,j=1,i\ne j}^{3}\Gamma_{e}\left(z_{1}^{(i)}(z_{1}^{(j)})^{-1}\right)\Gamma_{e}\left(z_{2}^{(i)}(z_{2}^{(j)})^{-1}\right)}
\ee
we can compare anomalies as we did for $N=2,\ k=3$, this comparison is displayed in table \ref{T:IT anomalies N=3 k=2},
\begin{table}[hbt]
	\begin{center}
		\begin{tabular}{ | c || p{9cm} | p{3cm} | c | }
  		\hline			
  		 & \ \ \ \ \ \ \ \ \ \ \ \ \ \ \ \ 4d interacting trinion & \ \ \ \ \ \ \ 6d prediction & $A_{IT}$\\ \hline\hline
  		$TrR'$ & $2[8\times6+(\frac12-1)4\times9\times 4]-2[(\frac{3}{2}-1)4 + (3-1) +3\times4+8\times1 + (-\frac12-1)4 + (\frac{3}{4}-1)16+(\frac1{4}-1)16] + 3[2\times 8] $ & $ -\frac12(2^2-2)(3-1)\cdot(2\cdot 2-2) $ & $ -4 $ \\ \hline
  		$Trt$ & $2[(-1)54+(\frac{1}{2})4\times9\times4]-2[(\frac{3}{2}) 4 + (1) 1 +(-1)24 +(-\frac12) 4+(\frac{3}{4})16+(\frac1{4})16] + 3[2\cdot 3^2(-1)]$ & $-\frac12 (2 \cdot 2^2\cdot 3 \cdot 1)$ & $-12$ \\ \hline
  		$Tr\beta$ & $-2[(3)3+(-\frac{3}{2})6+(\frac{3}{2})8 +(-\frac12)6 +(-1)6]$ & $2\cdot3\left(-\frac12-\frac12\right)$ & $-6$ \\ \hline
  		$Tr\gamma$ & $-2[(-3)3+(\frac{3}{2})6+(-\frac{3}{2})8 +(1)6 +(\frac12)6]$ & $2\cdot3\left(\frac12-(-\frac12)\right)$ & $6$ \\
  		\hline 
		\end{tabular}
	\end{center}
	\caption{Comparison of linear anomalies of the genus two Riemann surface built out of two interacting trinions, and the 6d predictions. The rest of the anomalies match as well. The  Riemann surface fluxes are ${\cal F}=\left( b_i=(-\frac12,\frac12),c_j=(\frac12,-\frac12),e=1\right)$.}
	\label{T:IT anomalies N=3 k=2}
\end{table}
and again give us validation of our results.

\end{appendix}


\bibliographystyle{ytphys}
\bibliography{refs}

\end{document}